\documentclass[sn-mathphys,Numbered]{sn-jnl}

\usepackage{setspace}
\usepackage{graphicx}%
\usepackage{multirow}%
\usepackage{amsmath,amssymb,amsfonts}%
\usepackage{amsthm}%
\usepackage{mathrsfs}%
\usepackage[title]{appendix}%
\usepackage{xcolor}%
\usepackage{textcomp}%
\usepackage{manyfoot}%
\usepackage{booktabs}%
\usepackage{algorithm}%
\usepackage{algorithmicx}%
\usepackage{algpseudocode}%
\usepackage{listings}%
\usepackage{booktabs}\usepackage{graphicx}%
\usepackage{geometry}

\definecolor{darkolivegreen}{rgb}{0.33, 0.42, 0.18}
\definecolor{celestialblue}{rgb}{0.29, 0.59, 0.82}

\usepackage{thmtools, thm-restate}
\usepackage{dsfont}
\usepackage{mathtools}
\mathtoolsset{showonlyrefs=false}
\usepackage{subfigure}
\usepackage{color}

\theoremstyle{definition}

\numberwithin{equation}{section} 

\newcommand{\cS}{\ensuremath{\mathcal{S}}}

\newcommand{\bR}{\ensuremath{\mathbb{R}}}

\newcommand{\He}{\ensuremath{\text{He}}}
\newcommand{\Rey}{\ensuremath{\text{Re}}}
\newcommand{\Hed}{\ensuremath{\text{He}}}
\newcommand{\bu}{\ensuremath{u}}

\newcommand{\bw}{\ensuremath{w}}

\def\[{\left[}
\def\]{\right]}
\def\<{\langle}
\def\>{\rangle}
\def\({\left(}
\def\){\right)}
\def\[{\left [}
\def\]{\right]}
\def\({\left(}
\def\){\right)}

\newcommand{\dx}{\ensuremath{~\mathrm dx}}

\providecommand{\abs}[1]{\lvert#1\rvert}

\newcommand{\lastiteration}[1]{#1}
\newcommand{\fg}[1]{#1}

\begin{document}

\newgeometry{top=2cm, bottom=2cm, left=2cm, right=2cm}

\onehalfspacing

\title[Article Title]{A parametric study of the pipeline hammer phenomenon in plastic Bingham slurry flows using the finite element method}

\author*[1]{\fnm{Felipe} \sur{Galarce}}\email{felipe.galarce@pucv.cl}
\author[1]{\fnm{Francisco} \sur{Martinez}}\email{francisco.martinez@pucv.cl}
\equalcont{These authors contributed equally to this work.}

\affil*[1]{\orgdiv{School of Civil Engineering}, \orgname{Pontificia Universidad Católica de Valparaíso}, \orgaddress{\street{Avenida Brasil 2147}, \city{Valparaiso}, \postcode{2340000}, \country{Chile}}}

\abstract{We conduct a numerical study of the transient phenomenon in pipelines transporting plastic Bingham slurry flows, using a lowest-order finite element method (FEM). While most pipeline hammer studies focus on Newtonian fluids, the transient dynamics in Bingham fluids remains elusive and poorly afforded, despite their significant industrial impact, particularly in mining. A detailed parametric study assesses the effects of the slurry yield stress and the valve closure times on both pressure and velocity distributions along the pipeline, using an adaptive friction model to account for turbulent slurries. Results reveal that yield stress enhances flow resistance and accelerates pressure peak attenuation, underscoring the damping role of Bingham rheology compared to Newtonian flows. These insights emphasize the need for advanced FEM-based schemes in non-Newtonian shockwave modeling, with implications for industrial pipeline design and operational safety.}


\maketitle

\section{Introduction}
\label{sec:intro}

Water-hammer is a common and critical phenomenon in many hydro-transport systems around the world. This problem is often caused by sudden operations in transport systems, such as a sudden closing or opening of flow control valves, pump or turbines power outages, or the pipeline ruptures, to name a few examples \cite{Dan_hammer2024, Qiaping2023, wylie1978fluid}. These events can generate intense pressure shockwaves traveling along the pipeline at very high speeds, covering long distances in quite short times. This feature gives very short time for a mechanical response to the pipeline walls, resulting in a significant increase in flow's pressure and, in extreme cases, approaching to the limiting mechanical strength of the pipeline. Then, this is a significant challenge for the operation of pipelines including system failure, equipment damage and operational inefficiencies, among other \cite{chaudhry2014}.

The hydraulic modeling of water-hammer has been intensely studied for Newtonian fluids, particularly for water transportation systems \cite{ghidaoui2004, ghidaoui2005}. However, an opposite situation occurs when dealing with complex fluids where few studies can be found \cite{kodura2018, kodura2019, kodura2022, rocha2008, ihle2014}. Non-Newtonian fluids are frequently encountered in mining industry, exhibiting operational problems similar to those found in water pipelines. The transport of slurries and mineral concentrates (e.g., copper tailings) and in general, solid-liquid suspensions of higher concentrations are good examples of rheological complex fluids, with both commercial interest and environmental impact \cite{boger2013, abdul2012, younger2004, cacciuttolo2022}. 

Recent studies have addressed this phenomenon affording specific topics such as shear-thinning, shear-thickening or even thixotropy effects on shockwave propagation \cite{wahba2013non}, missing an overall analysis about the influence of physical-mechanical parameters on the transient evolution on these flows. The complexity of the modeling of solid-liquid flows is usually related to the rheological character of the flow, often exhibiting a mechanical behavior similar to the plastic Bingham model. This model is particularly present for volume concentrations up to $C_{V} \approx 30\%$ \cite{abulnaga2021}. A dramatic increase in this concentration could lead to hyperconcentrated regimes governed by nonlinear rheological models \cite{obrien1985,obrien1988,julien1991}. 

On the other hand, mining tailings are usually characterized by viscosity values ($\eta$) up to 100 times higher than the viscosity of water and the existence of yield stress ($\tau_y$) affecting the initial mobility of the mixture\cite{abulnaga2021, wasp1977}. If flow's pressure and inertia are not able to overcome $\tau_{y}$, no flow is possible. Then it is clear that viscosity ($\eta$) and yield stress could lead to different dynamics with respect to classical water-hammer phenomenon, affecting the magnitude of the pressure wave, the shockwave speed and the attenuation dynamics \cite{mitishita2018}.

Current state of the art on the water-hammer numerical modeling includes several approaches, ranging on different trades-off between computational cost and accuracy. First, computational fluid dynamics (CFD) offer detailed simulations of transient evolution by directly solving the Navier-Stokes equations \cite{PHAM2021121099}. CFD models capture complex transient characteristics including shock waves, column separation and fluid-structure interaction. Advanced CFD tools incorporate adaptive mesh refinement, multi-phase flow modeling, and coupling with structural analysis to predict the response of piping systems and components accurately \cite{DARBHAMULLA2024106283}. Second, hybrid approaches combine the detailed flow dynamics captured by CFD codes with the computational efficiency of 1D models \cite{WANG20171}. 

These models are useful for analyzing systems where local three-dimensional effects are significant, although the overall behaviour of the event can be approximated as 1D \cite{oliveira2015}. Hybrid models enable detailed analysis of specific components (e.g., pump or valve regions) while efficiently analyzing the broader system. Finally, machine learning and artificial intelligence (AI) simulation techniques is gaining growing interest to predict and mitigate water hammer effects \cite{hammerIA}. These approaches can identify patterns from historical data, simulate various operational scenarios, and recommend preventive or mitigation strategies. Machine learning models can complement traditional simulation methods by providing insights derived from large datasets that are impractical to analyze with conventional methods \cite{Brunton_Kutz_2019}. 

\subsection{One dimensional models}

One-dimensional (1D) models are among the most popular schemes for water-hammer analysis. Often based on the method of characteristics (MOC) (e.g. \cite{wylie1978fluid, oliveira2015}), they can efficiently simulate shockwave propagation in pipes accounting for wall shear stress and friction, simplified flow-structure interaction models, presence of bifurcations or valve closure models, among others. Recent progress have improved the accuracy and computational cost of these models allowing for the simulation of large hydraulic systems. This approach becomes interesting when dealing with mining suspensions, considering the difficulty of solving the coupled motion equations associated to the solid and liquid phases \cite{fortier1967,adams2023}, but also because of the dynamical relations involved on the interaction between both phases require previous empirical validation which is not always available. The continuum modeling of pipeline hammer incorporating a rheological model emerges then as a simple and low-cost alternative for predicting pressure impacts in a pipe. 

In this context, this research aims to address the transient dynamics when dealing with plastic Bingham flows in horizontal, single diameter elastic-walled pipelines. Recent works have numerically afforded this case, studying the pressure waves generated after a sudden valve closure in a pipeline with Bingham fluids flows \cite{oliveira2010, oliveira2012, oliveira2015}, but also in power-law fluids \cite{santos2023}. Although these studies have reached significant progress, there are still several gaps that need to be addressed. By one hand, a better understanding on the influence of rheological parameters (e.g., $\eta,\tau_y$) on the propagation of pressure waves, changes in flow friction, water-hammer attenuation is required. On the other hand, the effects induced by gradual operations of some flow control instruments (e.g., valves), is also needed. This kind of operations obviously affect the decay dynamics of flow's velocity and pressure, the most important variables in the pipelines operation.

In our case, a methodological difference concerning the spatial discretization of the governing dynamics of the system was considered. The gold standard has relied on the method of characteristics \cite{chaudhry2014, oliveira2015} with a structured 2D grid for space and time. Taking into account the consolidated computational power of the last decades, we consider instead a finite element approach which allows one not only to solve the classical conservation equations, but also to extend them to several other scenarios. 

\fg{While the classical Method of Characteristics (MOC) has long served as the gold standard, particularly in conjunction with structured 2D grids for space and time \cite{chaudhry2014, oliveira2015}, its applicability is inherently limited to first-order hyperbolic PDEs, where nonlinearities appear only in the unknowns and not in their derivatives. This limitation becomes significant when addressing higher-order or fully nonlinear dynamics, such as those arising from more advanced viscous or constitutive models, where the characteristic structure may break down and numerical artifacts can appear without convergence guarantees. In contrast, the finite element method (FEM) provides a more general and robust framework that naturally accommodates such complex formulations. By allowing flexible discretization in both space and time and supporting weak formulations of arbitrary order, FEM enables the extension of our modeling approach beyond classical conservation laws to a broader class of non-Newtonian and nonlinear flow regimes. For example, extending the shear model using the admittedly simple power-law approach \cite{Abugattas2020} results in a second-order nonlinear PDE, for which characteristic curves cannot be consistently defined.} \lastiteration{Similarly, elasto-viscoplastic behavior can be analyzed with nonlinear models such as those explored in \cite{MONTEIROANDRADE2023104391}, where additional transport equations are incorporated into the overall dynamics.}

This paper is organized as follows: first, the governing equations for non-Newtonian transient phenomenon are introduced, showing its dimensionless form and the computation of Bingham friction factors. Then, the numerical scheme used to handle the equations in time and space is also described. Next, we show numerical experiments we have conducted to: first, test space and time convergence for the FEM scheme we use, second, to simulate a realistic copper slurry, third, to test out several yield stress configurations and asses their impact on the overall dynamics, and fourth, to asses the valve closure time impact on the pressure wave evolution and peak values. 

\section{Governing Equations}

Hydraulic transient equations for pipes are widely covered in literature, particularly for water transport. However, for non-Newtonian fluids like Bingham fluids or other complex rheologies, the available literature is much scarcer. In this context, we use a standard formulation for slurry flows \cite{oliveira2015}, used to model a pipe hammer with Bingham fluids in laminar systems of constant internal diameter and constant upstream head pressure. We include a friction model accounting for the laminar and turbulent hydrodynamic character of the flow, several alternatives for the valve closure model, and also the convective effects contained in the term $\bu \partial \bu / \partial x$, usually neglected in this kind of simulations. 

\begin{figure}[!htbp]
\centering
\includegraphics[width=0.8\textwidth]{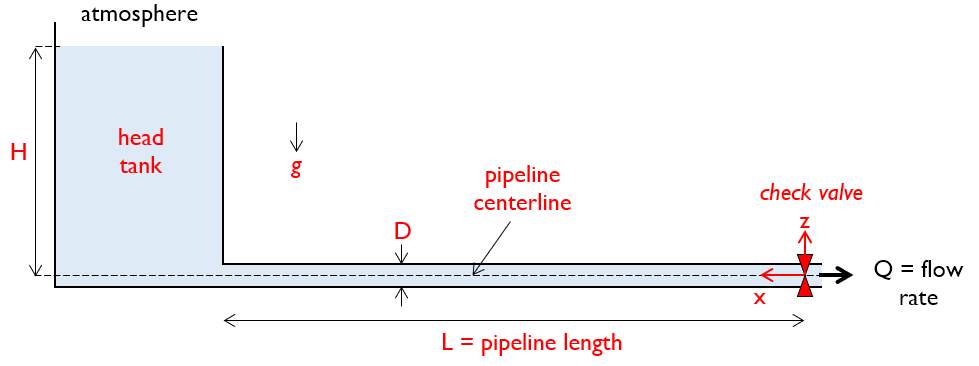}
\caption{Scheme used for pipeline hammer analysis. The main flow and pipe variables are defined. The head tank is open to atmosphere and flow is one dimensional for all purposes.}
\label{fig:setup}
\end{figure}

\fg{We model the velocity $u:[0,L] \rightarrow \bR$ and pressure fields $p:[0,L] \rightarrow \bR$ over the real interval (the pipe) $[0,L]$}. The wave propagation after the closure of a control valve at the pipeline downstream section, is modeled as depicted in Figure~\ref{fig:setup}. In this setup, an upstream head reservoir of height $H$ is connected to the pipe of inner diameter $D$ and total length $L$. A control valve is located at the downstream section of the system, which can close instantaneously ($T_c=0$) or gradually ($T_c>0$) depending on the operating conditions, where $T_c$ is the closure time. In this context, the one-dimensional (1D) averaged momentum and mass balance equations \cite{chaudhry2014, oliveira2015} read:
\begin{equation}
\left\lbrace
    \begin{aligned}
        \frac{\partial p}{\partial t} + \rho_m c^2 \frac{\partial \bu}{\partial x} &= 0 & \text{ in } (x,t) \in [0,L] \times [0,T], \\
        \rho_m  \left( \frac{\partial \bu}{\partial t}  + \bu \frac{\partial \bu}{\partial x} \right) + \frac{\partial p}{\partial x} + \cS(\bu) &= 0 & \text{ in } (x,t) \in [0,L]  \times [0,T], \\
        \bu &= \frac{16}{f(\widetilde{\Rey}, \He) \widetilde{\Rey}} g(t) & \text{ on } x = L,~t \geq 0, \\
        p &= P_o := \rho_m g H & \text{ on } x = 0,~t \geq 0,  \\
        p &= P_o \left( 1 - \frac{x}{L} \right) & \text{ at } t =0,  ~x \in [0,L],\\
        \bu &= \frac{16}{f(\widetilde{\Rey}, \He) \widetilde{\Rey}} U_{o}   & \text{ at } t =0, ~x \in [0,L],
    \end{aligned}        
\right.
\label{eq1}
\end{equation}
where  \fg{$f(\Rey, \He)$ is a non-constant friction factor, i.e., it varies in space and time with the local velocity, and $\rho_m$ stands for the mixture density, as it will be explained later in Section \ref{sec:wave}}. The parameters $P_o,U_o$ stand for characteristics pressure and velocity scales, respectively. Namely, we set $P_o = \rho_m g H$ and $U_o$ as the stationary laminar flow mean velocity before the valve closure:
\begin{equation}
U_o :=  \frac{P_o D^2}{32 L \eta},
\end{equation}
where $\eta$ is the fluid Bingham viscosity. In other words, we set $U_o$ as the Hagen-Poiseuille formula for average flow velocity in a circular pipe.

\begin{figure}
    \centering
    \includegraphics[width=0.45\linewidth]{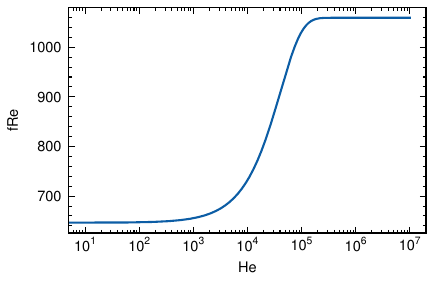}
    \includegraphics[width=0.45\linewidth]{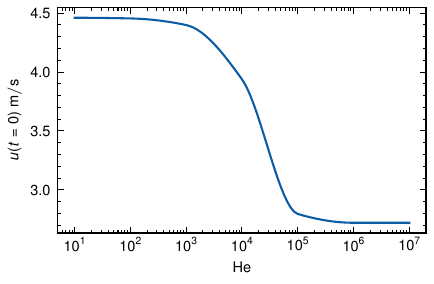}
    \caption{Initial conditions for the slurry system according to the parameter setup of a copper mixture, detailed in Tables \ref{tab:t1}, \ref{tab:t2} and \ref{tab:t3}}
    \label{fig:ini_condition}
\end{figure}

In addition, $\widetilde{\Rey}= \frac{\rho U_o D }{\eta}$ stands for a reference (and constant) Reynolds number computed with the characteristic velocity. On the other hand,  the \fg{instantaneous} Reynolds ($\Rey$) and Hedstrom ($\He$) numbers are introduced as follows \cite{abulnaga2021}: 

\begin{equation}
\begin{aligned}
\Rey &= \frac{\rho D u}{\eta},
\\
\text{He} &= \frac{\rho_m D^2 \tau_y}{\eta^2},
\end{aligned}
\label{eq4}
\end{equation}

where $\tau_y$ is the fluid yield stress. It's worth noting that the system initial condition will heavily depend on both, the Reynolds and the Hedstrom number. For turbulent regimes, higher friction factors will be computed, thus leading to $16/(f \Rey) \ll 1$. A similar phenomena will be depicted for large Hedstrom numbers, where high yield-stresses will naturally lead a larger friction with the walls. In Figure \ref{fig:ini_condition}, we show the starting conditions for friction and velocity of the system, for a range of Hedstrom numbers, using the parameter setup from \fg{Tables \ref{tab:t1}, \ref{tab:t2} and \ref{tab:t3}, for a 30\% concentration copper slurry}. The reader should note that the friction factor is a non-constant parameter in the simulation, thus increasing the overall numerical complexity as it is necessary to recompute the factor $f \Rey$ at each time-step, in each mesh node.

\fg{The reader will notice in the following sections that some simulations extend beyond the laminar regime. Nevertheless, we continue to use $U_0$ as a rough approximation for the initial 1D velocity field. To account for the non-Newtonian behavior of the fluid, we include a corrector term—sometimes referred to as \textsl{conductance} \cite{oliveira2012}—given by $16/(f \widetilde{\Rey})$. Regarding the pressure field, $P_0$ is imposed as a boundary condition at the head tank, located at $x = 0$ (see Figure \ref{fig:setup}), while the pressure distribution along the pipeline follows a linear steady-state profile consistent with a fully developed Poiseuille pipe flow, given by $p(x) = P_0 (1 - x/L)$. Finally, the valve closure at the downstream end ($x = L$) is modeled through a time-dependent function $g(t)$.}





The parameter $c$ represents the shockwave celerity. These scales will be used later on to formulate the dimensionless form of the governing equations. In addition, an isothermic assumption is considered for the fluid compressibility leading to a constant value for the pressure wave speed $c$. Therefore, the equations describe a fluid flow with mixture density $\rho_m$, axial velocity $\bu$, pressure $p$ and a reaction term denoted by $\cS(\bu)$. This last term models the wall shear stress effect on the overall motion. Using the Fanning friction factor $f$ \cite{white1991}, the cross-section velocity $u$ and pipe diameter $D$ we can write $\cS(\bu)$ as follows (\cite{chaudhry2014}, chapter 2):
\begin{equation}
\cS(\bu) = \frac{2 f(\Rey,\He) \rho_m |\bu| \bu }{D},
\label{eq2}
\end{equation}
\fg{which is a quasi-steady approach. This represents a simplified formulation of the wall shear stress model \cite{ghidaoui2005, duan2020_waterUWSS, PLOURABOUE2024237}, as unsteady velocity profiles have been shown to differ significantly from those in the steady-state regime \cite{riasi2009}, regardless of the fluid rheology \cite{wahba2013non}. \lastiteration{For example, in \cite{URBANOWICZ2023117848}, a comprehensive set of analytic expressions for unsteady friction losses is studied in the Newtonian scenario. Similarly, in \cite{filho2023} a comparative study involving six different models assesses them not only based on how they reproduce the wave propagation, but also under an entropy-based metric. In \cite{urbanoWater}, a convolution-based model is used for the friction losses, as is usually done with damping models where the equations are more easily handled in the frequency domain.}}

\lastiteration{Furthermore, recent studies \cite{XU2025123779} have integrated unsteady friction losses in contemporary optimization problems where the governing dynamics are introduced in a cost function where the state is approximated with a neural network \cite{RAISSI2019686}. This approach with physics-informed neural networks (PINNs) relates to a class of inverse problems where the physical information of the phenomena can be entangled with sensor data \cite{MPPY2015, galarce2023bias}, which is another future branch to be studied by the authors.}

The quasi-steady approach is broadly used by the community and is also present in other computer codes \cite{urbanowicz_2023, ELKHATIB2022139}, so it is still an open subject to be tackled in future works. We also remark that $\cS(\bu)$ will be computed locally in time and space, meaning that the friction factor will vary as a function of the fluid Reynolds number and the fluid Hedstrom number.

The fluid rheology is implicit in this system through the Fanning friction coefficient. For instance, for Bingham flows the coefficient $f \Rey$ can be readily estimated from the following relations~\cite{SWAMEE2011178, darby1981, darby1992}:

\begin{equation}
f(\Rey, \He) \Rey = \left\lbrace
\begin{aligned}
16 + \frac{10.67 + 0.1414 \left(\frac{\Hed}{\Rey}\right)^{1.143}}{1 + 0.0149 \left(\frac{\Hed}{\Rey}\right)^{1.16}} \left(\frac{\Hed}{4 \Rey} \right), \quad & \Rey < \Rey_c, \\
\Rey \left( \frac{ 10^{a} }{4 \Rey^{0.193} } \right), \quad & \Rey \geq \Rey_c,
\end{aligned}
\right.
\label{eq:friction}
\end{equation}
where $a = -1.47 (1 + 0.146 \exp(-2.9\cdot 10^{-5} \He))$ is an empirical coefficient \cite{abulnaga2021}, \fg{and $\Rey_c$ is a critical Reynolds number that sets the transition to turbulent regime. The calculation of $\Rey_c$ for Bingham flows will be discussed in Section \ref{sec:transition}}. Notice that these relations recover the Newtonian behavior when $\text{He}=0$. It should also be noted that wall roughness does not appear explicitly in the expression for the turbulent regime, unlike in water pipelines where the Colebrook–White formula is typically used. This is because such effects are already implicitly embedded in the empirical coefficients involved in the relationship, namely $a$ and the parameter 0.193.

\subsection{Shockwave propagation in homogeneous mixtures}
\label{sec:wave}

Among the assumptions made in this study, we will consider that the fluid is an homogeneous mixture, that is, a fluid whose solid concentration and velocity distribution can be considered practically uniform across the pipe diameter \cite{abulnaga2021, wasp1977}. The density of an homogeneous mixture ($\rho_{m}$) is defined as \cite{wasp1977}:
\begin{equation}
\rho_{m} = \rho_{s}C_{V} + \rho_{f}(1-C_{V})
\label{eqR}
\end{equation}
where $\rho_{s}, \rho_{f}$ denotes the solid and fluid densities, respectively and $C_{V}$ the solid volumetric concentration. Theoretical estimations of the pressure surge $\Delta P$ caused by a sudden change in the velocity regime in a pipeline is a longstanding engineering challenge. Joukowski was the first to determine the pressure rise generated in a rigid-walled pipe transporting a Newtonian liquid of density $\rho_m$ \cite{tijsseling2007}, that is: 

\begin{equation}
\Delta {P} = \rho_m c U_{o},
\end{equation}
\label{eq70}
where $U_{o}$ is the initial mean velocity just before the close of the valve, $c$ is the shockwave speed. Since $c \gg U_{o}$, the wave speed is usually considered independent on the initial flow velocity. A general expression for the shockwave speed valid for incompressible, Newtonian liquids in single-diameter tubes, can be written as follows: 
\begin{equation}
c = \sqrt{ \frac{\frac{K_{f}}{\rho_{f}}}{1+\zeta} },
\end{equation}
where $K_{f}$ is the liquid bulk modulus. According to Young \cite{young1808}, for rigid walls pipelines we have $\zeta = 0$, and Korteweg showed \cite{korteweg1878a, korteweg1878b} that $\zeta = 1 + \frac{K_{f}}{E}\frac{D}{e}$ for elastic walls tubes, where $E$ is the pipe elastic modulus, $D$ the inner diameter of the tube and $e$ the wall thickness. Other expressions considering pipe constraints can be found in \cite{chaudhry2014, streeter1993}. The previous formulas can be adapted for estimating the wave speed celerity $c_{m}$ for homogeneous and heterogeneous solid-liquid mixtures. In the former, the expression reported by Young can be simply adapted as follows:
\begin{equation}
c = \sqrt{\frac{K_{m}}{\rho_{m}}},
\label{eqA}
\end{equation}
where $K_{m}, \rho_{m}$ denotes the bulk modulus and mixture density, respectively. A way for estimating $K_{m}$ is similar to Eq. (\ref{eqR}), that is $K_{m} = K_{s}C_{V} + K_{f}(1-C_{V})$, where $K_{s}$ is the bulk modulus of the slurry solid phase. The Korteweg law can also be written for solid-liquid mixtures as follows \cite{kodura2022}:
\begin{equation}
c = \sqrt{\frac{\frac{K_{m}}{\rho_{m}}}{1 + \frac{K_{m}}{E}\frac{D}{e}}}.
\label{eqB}
\end{equation}

Finally, another expression also valid for homogeneous mixtures read \cite{thorley1979, han1998}: 
\begin{equation}
c = \sqrt{\frac{\frac{K_{f}}{\rho_{m}}}{1 -C_{V} + \frac{K_{f}}{K_{s}}C_{V} + \frac{D}{e}\frac{K_{f}}{E} }}.
\label{eqC}
\end{equation}

In our numerical experiments, we will conduct simulations with this last expression, as it will be made clear in section \ref{sec:copper_slurry}.

\subsection{The transition velocity for Bingham flows}
\label{sec:transition}

The parameter $\Rey_c = \rho u_{c}D / \eta$ from \eqref{eq:friction}, can be considered as a critical flow Reynolds number indicating the laminar-turbulent separation regime. Such separation occurs at the velocity $u_{c}$, also called \textit{transition velocity} \cite{wasp1977}. When $u<u_{c}$ the flow regime is laminar, and turbulent when $u>u_{c}$. This critical parameter can be determined by following the method proposed by Hanks and Pratt \cite{hanks1967}:

\begin{equation}
\Rey_{c} = \frac{\He}{X_{c}} \left( 1 - \frac{4}{3}X_{c} + \frac{1}{3}X_{c}^{4}\right),
\label{2.6}
\end{equation}
\fg{where $X_{c} = \tau_{y}/(\tau_{w})_{c}$ and $(\tau_{w})_{c}$. Just at the laminar-transition point Hanks and Pratt \cite{hanks1967} demonstrated that Hedstrom number could be correlated with the critical Reynolds number, fixing the onset of turbulence. This correlation can be written as follows:}

\begin{equation}
\He = 16800 \frac{X_{c}}{(1-X_{c})^{3}}
\label{eq:X_c}
\end{equation}





This means we can compute the critical velocity and therefore the transition number $\text{Re}_c$ fixing a Hedstrom number, thus allowing us to study several rheological behaviors of the flow. Remark as well that $X_c$ can be physically interpreted as a Bingham number $\text{Bi} = \tau_y / (\tau_w)_c$ for a fully developed Bingham flow, which is the only regime in which this relation is valid. \fg{Just for reference, for the Newtonian case, the threshold value is $\Rey_c = 2,100$ \cite{assefa2015}, but in a slurry this threshold can be higher.}



\subsection{Dimensionless equations}

The system given in Eq.(\ref{eq1}) can be expressed in dimensionless form by introducing relevant characteristic scales for the velocity
($U_o$), pressure ($P_o$), position ($L_o$) and time ($T_o$) \cite{oliveira2010, oliveira2012, oliveira2015}. The scales for velocity and pressure were already introduced in the system \eqref{eq1}. Exploiting the simplicity of the experimental setup, the time scale is chosen as $T_{o} = L_o/c$, for $L_{o} = L$. These definitions leads to the unknowns $x^{*},t^{*},u^{*},p^{*}$ defined as:

\begin{subequations}
\begin{align}
x^{*} &= \frac{x}{L_{o}} = \frac{x}{L}
\\
t^{*} &= \frac{t}{T_{o}} = \frac{tc}{L}
\\
p^{*} &= \frac{p}{P_o} = \frac{p}{\rho g H} \\
\bu^{*} &= \frac{\bu}{U_o} = \bu \frac{32}{P_o} \frac{\eta L}{D^2}
\end{align}
\label{eq5}
\end{subequations}
which leads to the following system of equations:

\begin{equation}
\left\lbrace
    \begin{aligned}
        \frac{\partial p^*}{\partial t^*} + \lambda_2 \frac{\partial \bu^*}{\partial x} &= 0 & \text{ in } (x^*,t^*) \in [0,1] \times [0,T^*], \\
        \lambda_2 \frac{\partial \bu^*}{\partial t^*}  + \lambda_1 \bu^* \frac{\partial \bu^*}{\partial x^*} + \frac{\partial p^*}{\partial x^*} + \frac{f(\Rey, \He) \Rey}{16} \bu^* &= 0 & \text{ in } (x^*,t^*) \in [0,1]  \times [0,T^*], \\
        \bu^* &= \frac{16}{f(\widetilde{\Rey}, \He) \widetilde{\Rey}} g^*(t) & \text{ on } x^* = 1,~t^* \geq 0, \\
        p^* &= 1 & \text{ on } x = 0,~t^* \geq 0,  \\
        p^* &= 1 - x^* & \text{ at } t^* =0,  ~x^* \in [0,1],\\
        \bu^* &= \frac{16}{f(\widetilde{\Rey}, \He) \widetilde{\Rey}}  & \text{ at } t^*=0, ~x^* \in [0,1],
    \end{aligned}
\right.
\label{eq:governing_scaled}
\end{equation}
and where we have introduced,

\begin{subequations}
\begin{align}
\lambda_1 &= \frac{\delta}{32} \widetilde{\Rey} \\
\lambda_2 &= \frac{\delta}{32} \widetilde{\Rey} ~\widetilde{\text{Ma}}^{-1}
\end{align}
\label{eq7}
\end{subequations}
where $\widetilde{\text{Ma}} = U_o/c$ is a reference Mach number defined with the stationary velocity $U_o$, and $\delta = D/L$ is an aspect ratio for the pipeline geometry.

\subsection{A lowest-order finite element formulation for slurry flows}
\label{sec:methods}


Among the alternatives to perform to numerically solve in space and time the equations \eqref{eq:governing_scaled} we can name the MOC \cite{oliveira2015}, finite volumes \cite{KERGER20112030}, and the finite element method (FEM, \cite{fem_bible} \cite{fem_ern}). We will use the latter to deal with space discretization and implicit finite differences for time. Let us consider a one-dimensional working domain, a straight line represented by the real interval $\Omega = [0, L]$ ($L \in \mathbb{R}$). In order to properly expose the numerical setup, it is required to formalize the functions $u^*:[0, L] \rightarrow \mathbb{R}$ and $p^*:[0, L] \rightarrow \mathbb{R}$ as an element of a \fg{product} Hilbert space $\mathcal{H} = H^1([0,L]) \times L^2([0,L])$, where $L^2 = \{ v : \Omega \rightarrow \Omega : \int_\Omega \abs{v}^2 \dx < \infty \}$, and $H^1(\Omega)$ stands for the smaller space of smoother functions $H^1 = \{ v \in L^2(\Omega); ~\partial v / \partial x \in L^2(\Omega)\}$.

\fg{In this setting, the problem can be recast in variational form, reducing the task to finding a pair of functions $(\bu^{*,n+1}, p^{*,n+1}) \in \mathcal{H}$ such that:}

\begin{equation}
\begin{aligned}
\frac{1}{\Delta t^*} \int_{\Omega} p^{*,n+1} q \dx + \lambda_2 \int_\Omega \frac{\partial \bu^{*,n+1}}{\partial x^*} q \dx + \frac{\lambda_2}{\Delta t^*} \int_\Omega \bu^{*,n+1} \bw \dx + \int_\Omega \frac{\partial p^{*,n+1}}{\partial x^*} \bw \dx + 
\\
+ \lambda_1 \int_{\Omega} u^* \frac{\partial u^*}{\partial x^*} \bw \dx + \frac{f \Rey}{16} \int_\Omega \bu^{*,n+1} \bw \dx = \frac{1}{\Delta t^*} \int_{\Omega} p^{*,n} q \dx + \frac{\lambda_2}{\Delta t^*} \int_\Omega \bu^{*,n} \bw \dx,
\end{aligned}
\label{eq:weakForm}
\end{equation}
for all $q \in L^2(\Omega)$ and $w \in H^1(\Omega)$. We have introduced an implicit Euler scheme for time differentiation, i.e., the first-order approximations:

\begin{subequations}
\begin{align}
 \frac{\partial \bu^*}{\partial t^*}  = \frac{\bu^{*,n+1} - \bu^{*,n}}{\Delta t^*},
\\
 \frac{\partial p^*}{\partial t^*} = \frac{p^{*,n+1} - p^{*,n}}{\Delta t^*},
\end{align}
\label{eq:variational}
\end{subequations}
which means that we need to solve the system \eqref{eq:variational} consecutively to complete the simulation time with a uniform \fg{time} grid $[0,\Delta{t}, 2 \Delta{t}, \ldots]$. A traditional Picard method is used to deal with the system non-linearities at each time-step, which will have a fixed error tolerance over the $\ell^2$ norm for joint velocity and pressure of $10^{-3}$. Next, we recast the problem in finite dimension by means of the following standard Galerkin approach: 
\begin{equation}
\begin{aligned}
\bu^{*,n+1} = \sum_{i=1}^N \bu^{*,n+1}_i  \phi_i, \\
p^{*,n+1} = \sum_{i=1}^N p^{*,n+1}_i \phi_i,
\end{aligned}
\label{eq10}
\end{equation}
where $N$ is the number of nodes that discretize the working domain $\Omega$. To ease the reading, the same notation has been adopted for both the continuous and the discrete unknowns. The finite element method provides the solution of nodal coefficients $\bu^{*,n+1}_i$ and $p^{*,n+1}$ ($i=1,\ldots,N$). We choose the same discretization space for both, velocity and pressure, namely, the \fg{continuous} piece-wise linear (lowest-order) space $\mathbb{P}_1$ basis $\phi_1, \ldots, \phi_N$. Then, it is easily shown that this approximation leads to the following matrix system of $2N \times 2N$ equations:

\begin{equation}
    \begin{bmatrix}
    ( 1  + \Delta t^*\frac{f \Rey}{16}  ) M + \lambda_1 \Delta t^* C(u^{*,n})    \quad & \Delta t^* B  \\
    \lambda_2 \Delta t^* B                                           &  M  
    \end{bmatrix}
    \begin{bmatrix}
    \bu^{*,n+1} \\
    p^{*,n+1}
    \end{bmatrix}
    =
    \begin{bmatrix}
     \lambda_2 M \bu^{*,n} \\
     M p^{*,n}
    \end{bmatrix}
    \label{eq:matrix_fem}
\end{equation}
where $M \in \bR^{N \times N}$ stands for a mass matrix with entries $M_{ij} = \int_{\Omega} \phi_i \phi_j \dx$, the matrix $B\in \bR^{N \times N}$ for the coupling terms, with entries $B_{ij} = \int_{\Omega} \phi_i \frac{\partial \phi_j}{\partial Z} \dx$, and the convection matrix $C\in \bR^{N \times N}$ with entries $C_{ij} = \int_\Omega u^* \frac{\partial \phi_i}{\partial x^*} \phi_j \dx$. The integrals are computed with a standard Gaussian quadrature rule of three points per finite element. In addition, a classical fixed-point root-finder is utilized to solve for $X_c$ in equation \eqref{eq:X_c} for a given input Hedstrom number, thus allowing us to compute the yield stress and the nondimensional constants $\lambda_1$ and $\lambda_2$.

The assembly of the matrices and the solver is handled with the in-house software MAD \cite{galarceThesis} (used already in several CFD works with finite elements, e.g. \cite{GLM2021, galarce-etal-2023, galarce2023bias, GLM2021_2, galarcePachecoBVS}), for finite elements and data assimilation, which supports CPU parallelization for optimal speed-up thanks to the underlying linear algebra library PETSc\cite{petsc}. For this study, we use an AMD Epyc processor with 48 physical cores. We use the MUltifrontal Massively Parallel sparse direct Solver (MUMPs, \cite{AMESTOY2000501, mumps}) to compute the solutions of \eqref{eq:matrix_fem}, . 

The reader may note the absence of stabilization techniques, as it is the gold standard for the discretization of the Navier-Stokes equations.  In particular, we remark on the importance of the mass storage component $(1/\Delta t^*) M$ in the second line of \eqref{eq:matrix_fem} as a regularizer that prevents the system of equations from getting close to singular. One should consider nonetheless that getting close to a divergence-free scenario may distort the condition number of the matrix in \eqref{eq:matrix_fem}. On the other hand, it would be easy to counter a divergence-free situation. There are several stabilization techniques that would take this situation under control, such as the Brezzi-Pitk\"aranta approach \cite{brezzi1984}, the streamline upwind Petrov-Galerkin one \cite{supg}, or variational multi-scale methods \cite{GONZALEZ20201009}. \fg{Additionally---we can speculate that---since we are not operating in regimes where the convective term is dominant, upwind stabilization strategies are not necessary.} The later will be explored in a forthcoming scientific study to test out unstable regimes ensuring numerical resolvability. In addition, the reader may notice that we follow a monolithic scheme for the problem unknowns. Other decoupling strategies for velocity and pressure for this application may be studied in forthcoming research \cite{overviewSolvingNSincomp}. 

\section{Results}
\label{sec2}

This section is devoted to study the numerical solutions of the transient slurry equations with convective effects. We first set up, in section \ref{sec:convergence} the necessary mesh and time step for the finite element scheme we employ so that we reach non-oscillatory or not excessively regularized solutions, as a convergence study. Next, in section \ref{sec:copper_slurry}, we inspect the numerical solution for a realistic parameter set that emulates copper-water mixture. We then do a parametric study of the solutions including the effect of the fluid yield stress (section \ref{sec:impact_tauy}) and the impact of the valve closure time (section \ref{sec:impact_valve}).

\subsection{Convergence study}
\label{sec:convergence}

\fg{Let us begin by testing the convergence of the solver in both space and time discretization. For this purpose, we solve the governing equations using the set of parameters shown in Tables \ref{tab:t1}, \ref{tab:t2}, and \ref{tab:t3}, but with $\tau_y = 0$, as it poses the most challenging numerical scenario. This choice corresponds to sharp time derivatives in the solution and the least dissipative case. Without sufficient refinement, such sharp gradients may lead to numerical oscillations that would otherwise be damped in more viscous regimes. This test is important to assess whether typical stabilization techniques, such as streamline-upwind schemes or modern variational multiscale methods \cite{GONZALEZ20201009}, are necessary.}

\fg{Physically, the most unfavorable scenario concerning the valve closure model, occurs when the closure time is shorter than the round-trip wave travel time along the pipe, i.e., $T_c \leq \frac{2L}{c}$. As will be shown in Section~\ref{sec:impact_valve}, the limiting case $T_c = 0$ is numerically challenging, as simulations become prone to oscillations, due to the presence of extremely sharp wave fronts.}

\begin{figure}[!htbp]
    \centering    
    \subfigure[Space convergence test]{
    \includegraphics[width=0.46\textwidth]{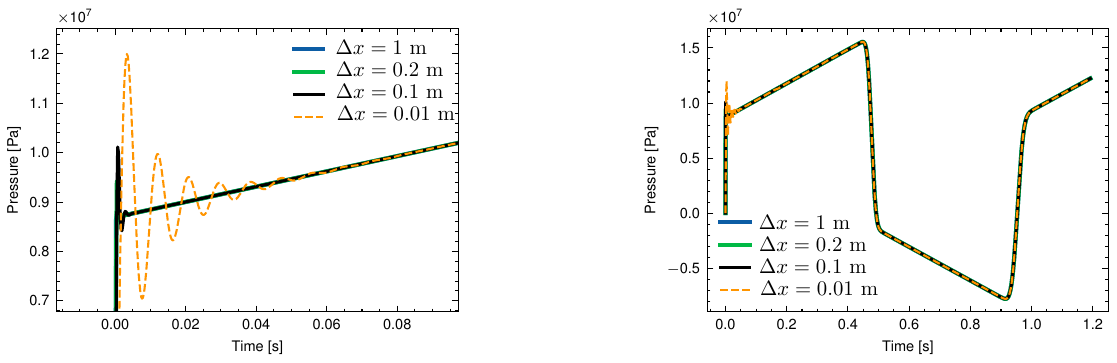}}
    \subfigure[Zoom-in and numerical oscillations]{\includegraphics[width=0.45\textwidth]{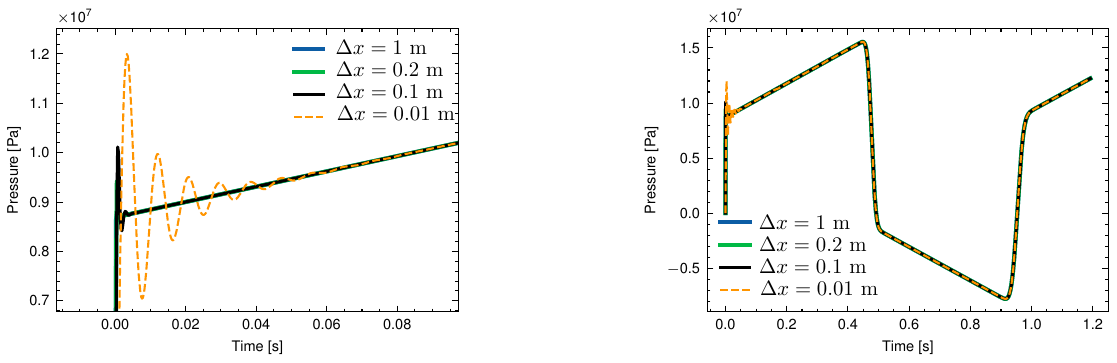}} 
    \caption{Mesh dependence test. As expected, the mesh refinement is critical after the instantaneous valve closure. This sets a threshold for the element size at $\Delta x = 0.2$ m \fg{($L=100 m$, with 1001 degrees of freedom).} \fg{The reader should notice that pressure magnitudes are relative to atmospheric values}}
    \label{fig:convergence_space}
\end{figure}

Thus, to analyze convergence we set a sudden closure for the valve, meaning that in the boundary conditions of \eqref{eq1} we choose \fg{$g(t)=0$ at $x=L$}. To study the mesh convergence, we narrow our attention to the pressure at a control point located at the valve. Consider a preliminary time step of $\sim 10^{-7}$ s and 4 meshes, with characteristic element sizes of $1$ m, $0.2$ m, $0.1$ m, and $0.01$ m. We observe, in Figures \ref{fig:convergence_space}.a and \ref{fig:convergence_space}.b, a mesh independent solution up to $\Delta x = 0.2$ m, so we choose this element size for all the forthcoming simulations. The critical jump is that of the wave onset, at the beginning of the simulation, so we zoom in the phenomena in Figure \ref{fig:convergence_space}. For every simulation, we check how the reservoir pressure is recovered after the flow stops, as it is physically \fg{expected for the final hydrostatic condition}.  

\begin{figure}[!htbp]
    \centering
    \subfigure[Time domain]{
    \includegraphics[width=0.96\textwidth]{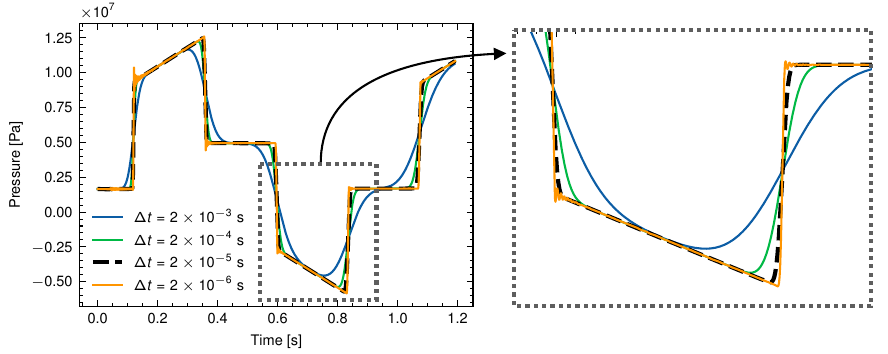}}
    \subfigure[Frecuency domain]{
    \includegraphics[width=0.45\textwidth]{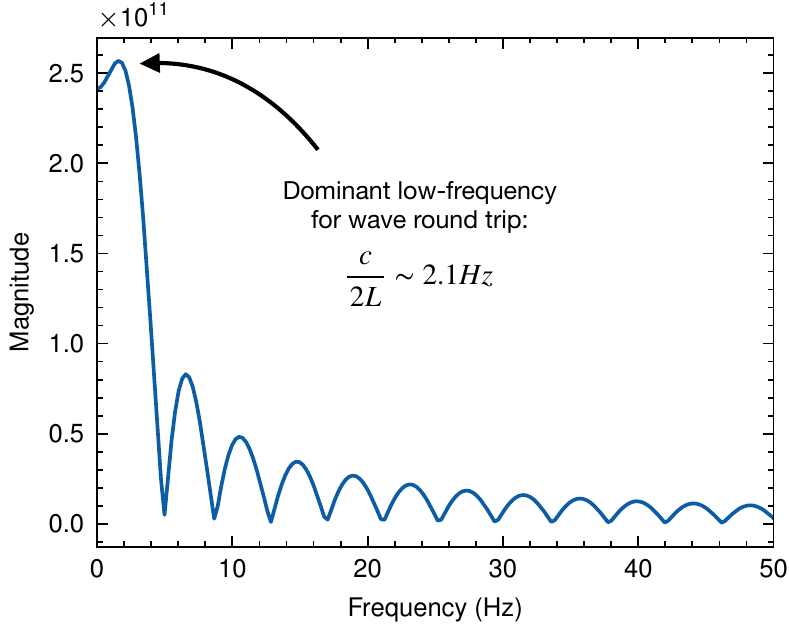}}
    \caption{Convergence test for Newtonian case $\text{He} = 0$. This benchmark shows that $\Delta t = 2 \times 10^{-5}$ s suffices to capture the hammer dynamics. \fg{In addition, the spectral content of the signal for the valve control point behaves as expected, with dominant frequencies as multiples of the round-trip wave}}
    \label{fig:fftComp}
\end{figure}

We proceed analyzing the time convergence using $\Delta x = 0.2$ m. Figure \ref{fig:fftComp}.a shows that $\Delta t \geq 2 \times 10^{-3}$ s leads to excessively smooth solutions. \fg{In addition, we analyze the frequency content of the pressure signal at the valve ($x = L$), as shown in Figure \ref{fig:fftComp}.b. We applied a Fast Fourier Transform (FFT) to a $0.48\ \mathrm{s}$ segment of the simulated time history—the duration required for the wave to travel to the reservoir and back. This time‐step selection is adequate: it captures the dominant low‐frequency component near $2.1\ \mathrm{Hz}$ and the subsequent peaks corresponding to the characteristic wave frequency $f = \frac{c}{2L}$.}

\fg{Due to the fact that we use a constant bulk modulus and omit cavitation, the simulations predict negative pressures well below the fluid's vapour pressure (as depicted in Figures \ref{fig:convergence_space} and \ref{fig:fftComp}). In practice, cavitation would prevent such large tensile stresses.}

\fg{We conclude that $\Delta t = 2 \times 10^{-5}$ s suffices to recover a coherent and physical solution, therefore we use it for all the simulations to come. In addition, we remark the absence of numerical viscosity achieved at this refinement level, which is a common issue in certain discretization schemes. It worth to mention that one could achieve equally good results with coarser meshes and time steps if we stabilize our finite element solver with classical methods such as the streamline-upwind Petrov-Galerkin approach \cite{BROOKS1982199}, or modern methods such as the variational multi-scale strategy \cite{RamonReview2017}. This is left as a future work for the authors.}

Concerning the solver performance---which in this test case involves the inversion of 50000 systems of equations with 1001 degrees of freedom (one each time step)---it takes a total simulation time of 66.34 seconds, using a single core of a 48 cores AMD Epyc processor.

\subsection{Simulation of a copper slurry}
\label{sec:copper_slurry}

In this section we will analyze the evolution of a hydraulic transient in a copper-water mixture pipeline system.  

\begin{table}[!htbp]
\caption{Physical and mechanical parameters for a copper slurry. Source: \cite{alejo2009}}
\label{tab:t1}
\begin{tabular}{@{}ccc@{}}
\toprule
Parameter   & Meaning                   & Value    \\ \midrule
$\rho_{s}$  & Solid density            & 8.9 g$\cdot$cm$^{-3}$ \\
$\rho_{f}$  & Liquid density (water)    & 1.0 g$\cdot$cm$^{-3}$ \\
$K_{f}$     & Liquid bulk modulus       & 2.1 GPa \\
$K_{s}$     & Solids bulk modulus       & 140 GPa \\
$E$         & Pipe elastic modulus (steel) & 200 GPa \\
$\eta$      & Slurry Bingham viscosity    & $\approx$ 0.03 Pa s   \\
$\tau_y$    & Slurry yield stress    & $\approx$ 26 Pa   \\
\end{tabular}
\end{table}

Table \ref{tab:t1} shows the physical and mechanical properties for the slurry flow considered in this study, Table \ref{tab:t2} the mixture density and bulk modulus and the expected theoretical values for the shockwave speed estimated according to the relations given by Eq.\ref{eqA} - Eq.\ref{eqC} for increasing values of $C_{V}$. We conduct our simulations using $C_V = 30 \%$. Table \ref{tab:t3} display the pipeline properties. \fg{The slurry physical properties lead to $\He = 1.02 \cdot 10^6$ \cite{abulnaga2021, alejo2009} and $u_c =$ 1.34 m$\cdot$s$^{-1}$, then $\Rey_c = 15,390$. For this numerical example the arbitrary value $U_{o} =$ 2.72 m/s was selected as the initial flow velocity. This choice ensures that the initial flow Reynolds number is $\Rey = 31,257 > \Rey_c = 15,390$, placing the system in the turbulent fully-suspended regime.} 


\begin{table}[!htbp]
\caption{Schockwave speeds ($c$) for a copper tailing according to different authors. In the last column, overpressure values ($\Delta {P}$) were included according to Joukowski formula $\Delta P = \rho_{m} c U_{o}$}
\label{tab:t2}
\begin{tabular}{@{}ccccccc@{}}
\toprule
Concentration & $\rho_{m}$ & $K_{m}$ & Eq.\ref{eqA} & Eq.\ref{eqB} & Eq.\ref{eqC} & $\Delta {P}$\\ 
$C_{V}$ & g$\cdot$cm$^{-3}$ & GPa & m/s & m/s & m/s & MPa\\ \midrule
10\%    & 1.79 & 15.9 & 2979 & 1944 & 1042 &  5.07 \\
15\%    & 2.19 & 22.8 & 3229 & 1885 & 966  &  5.75 \\
20\%    & 2.58 & 29.7 & 3392 & 1808 & 911  &  6.39 \\
25\%    & 2.98 & 36.6 & 3506 & 1730 & 870  &  7.05 \\
30\%    & 3.37 & 43.5 & 3592 & 1658 & 840  &  7.71 \\
\end{tabular}
\end{table}

Although in the range $C_{V}>30\%$ the mixture can still display a non-Newtonian behavior, a Bingham-like rheology will not be necessarily expected \cite{abulnaga2021}. For higher concentrations, an hyper-concentrated flow regime can be reached that is characterized by a different rheological law (see \textit{e.g.} \cite{julien1991}). In the same context, in the range $C_{V}<1\%$ a Newtonian-like regime is expected for the mixture, which can be addressed with usual methods \cite{wylie1978fluid, chaudhry2014}. 


\begin{table}[!htbp]
\caption{Parameter setting for the pipeline. The steel pipeline is class DN100 Sch.40s \cite{pipe}}
\label{tab:t3}
\begin{tabular}{@{}ccc@{}}
\toprule
Parameter   & Meaning              & Value    \\ \midrule
$L$         & Pipe length          & 200.0 m   \\
$H$         & Reservoir Head       & 100.0 m   \\
$D$         & Inner pipe diameter  & 102.3 mm  \\
$e$         & Wall thickness       & 6.0 mm    \\
$P_{A}$     & Allowable pressure   & 17.8 MPa   \\
\end{tabular}
\end{table}

In agreement with collected experience in engineering projects, results provided by Eqs. \eqref{eqA}-\eqref{eqB} are extremely large and thus unrealistic, \fg{and may need experimental verification in the context of copper slurry transients}. The wave speed values calculated from Eq.~\eqref{eqC} can be considered most representative of a real transient event, in agreement with Kodura et al. \cite{kodura2018, kodura2019, kodura2022}. Even more, it is expected that real values of the initial pressure surge are lower than theoretical estimations inspired on Joukowski's pressure scale, because of the lower shockwave speed (see e.g., \cite{kodura2018, kodura2022}). \fg{Thus, all our numerical simulations consider the wave speed computed from Eq.~\eqref{eqC}} 

\fg{With these elements, the Joukowsky's pressure rise can be easily estimated from Eq.\ref{eq70} as $\Delta {P} = \rho_{m} c U_{o}$. The last column of Table \ref{tab:t2} shows some estimations of $\Delta P$ by considering the reference velocity $U_{o}$. By one hand, the maximum pressure values along the pipeline $p_{max}(x)$ obtained from numerical simulations can be compared to the theoretical maximum pressure $\max_{0\leq x \leq L} (p(x,t=0) + \Delta P)$, where $p(x,t=0)$ is the steady-state pressure profile of the pipeline. On the other hand, $p_{max}(x)$ can also be compared to $P_{A}$, where $P_{A}$ is the \textit{allowable pressure}. As expected, a risky situation for pipeline safety occurs when $P_{max} \geq P_{A}$. This last parameter is provided by the pipeline factory and it depends on the wall material and schedule (see Table \ref{tab:t3}). In addition, the minimum pressures in the system, $P_{min}(x)$, can also cause problems in the pipeline. In this case, the risk situation will occur when $P_{min}(x) < P_v$, where $P_v$ is the vacuum pressure that induces the liquid-gas phase separation in the mixture. A typical value for this parameter is $P_v = -51.32$ Pa (using a gauge pressure scale).}

\begin{figure}[!htbp]
    \centering
    \subfigure[$u(t=0)$]{
    \includegraphics[height=4cm]{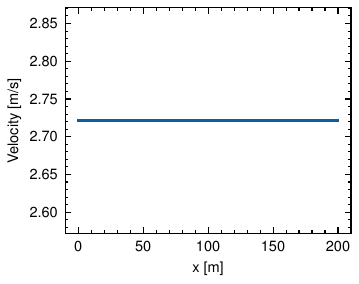}}
    \subfigure[$u(t=0.1 s)$]{
    \includegraphics[height=4cm]{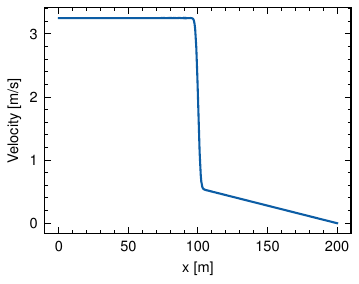}}
    \subfigure[$u(t=0.5 s)$]{
    \includegraphics[height=4cm]{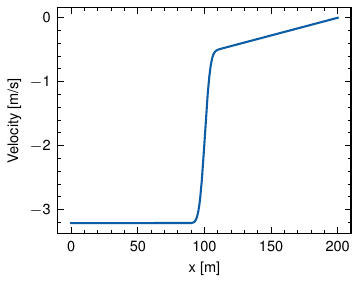}} \\
    \subfigure[$p(t=0)$]{
    \includegraphics[height=3.9cm]{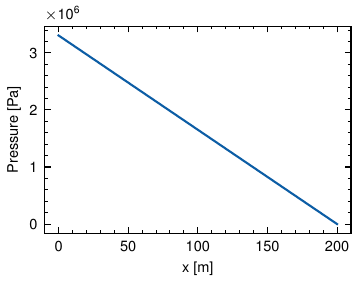}} 
    \subfigure[$p(t=0.1 s)$]{
    \includegraphics[height=3.9cm]{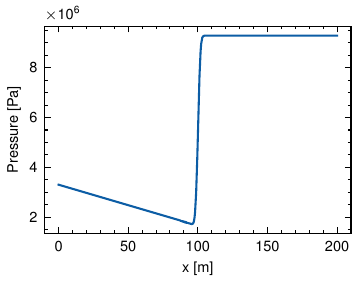}} 
    \subfigure[$p(t=0.5 s)$]{
    \includegraphics[height=3.9cm]{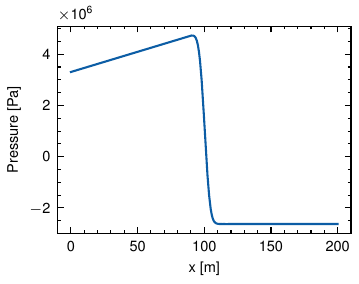}} \\   
    \caption{Numerical solution snapshots at first wave cycle}
    \label{fig:copper_solution_snap}
\end{figure}

In Figure \ref{fig:copper_solution_snap} we observe simulation snapshots during the phenomena onset after an instantaneous valve closure. In the figure, we show the system stationary initial condition, the over-pressure wave propagation, i.e. from the valve to the tank (at $t=0.1$ s) and the under-pressure, i.e., from the tank to the valve (at $t=0.5$ s).

\begin{figure}[!htbp]
    \centering
    \includegraphics[width=\textwidth]{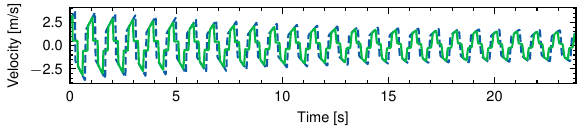}
    \caption{Velocity for copper slurry on 2 control points, at $x = 0$ (blue) and $x=L/2$ (green)}
    \label{fig:ctrlU}
\end{figure}

\begin{figure}[!htbp]
    \centering
    \includegraphics[width=\textwidth]{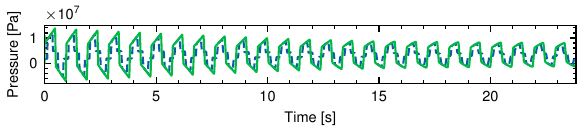}
    \caption{Pressure for copper slurry on 2 control points, at $x=L/2$ (blue) and $x=L$ (green)}
    \label{fig:ctrlP}
\end{figure}

In addition, we show the temporal evolution of the numerical solutions for control points located at the tank ($x=0$), the pipe half length ($x=L/2$), and the valve ($x=L$), at Figures \ref{fig:ctrlU} and \ref{fig:ctrlP}. We observe a rather slow friction dissipation and we check that the maximal numerical pressure reach a value of $13.86$ MPa, which turns out to be surprisingly close to the factory allowable pressure $P_A = 17.8$ MPa, but still ensuring a safe hammer wave propagation. The theoretical pressure rise discussed above, with more restricted assumptions, is lower than the numerical one, namely $11$ MPa, yet in accordance with the obtained value. 

\subsection{Impact of the fluid yield stress}
\label{sec:impact_tauy}

\begin{figure}[!htbp]
    \centering
    \includegraphics[width=0.6\textwidth]{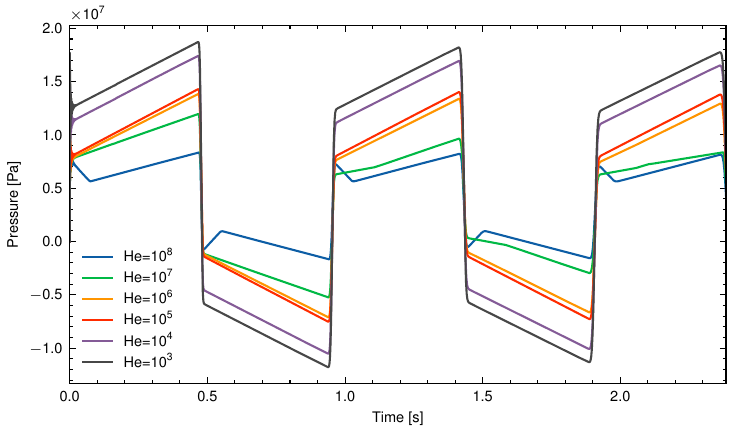} 
    \caption{Pressure evolution at valve control point, i.e. $x=L$.}
    \label{fig:up_vs_Bi1}
\end{figure}

\begin{figure}[!htbp]
    \centering
    \includegraphics[width=0.6\textwidth]{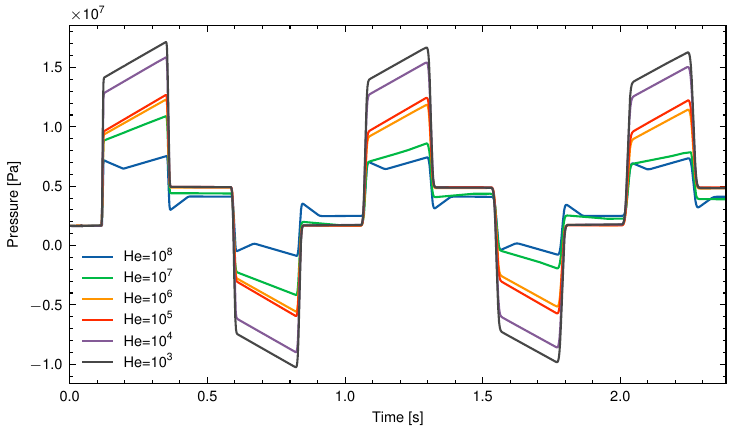} 
    \caption{Pressure evolution at half pipe control point, i.e. $x=L/2$.}
    \label{fig:up_vs_Bi2}
\end{figure}

A comparison for several yield stress magnitudes is depicted in Figures \ref{fig:up_vs_Bi1} and \ref{fig:up_vs_Bi2}. To convey this simulations, we have used the same parameter configuration as for the previous test case, but moving freely the Hedstrom number and thus recomputing the yield stress at each case. In these figures, we can see how larger Hedstrom numbers simulations depicts smaller peaks in over-pressure values, leading naturally to faster decay times. This is a natural an expected behavior, as it is intuitive that larger Hedstrom numbers will imply more difficulty to move the fluid in the first place so that the yield stress could be reached, thus leading to a damping effect seen in the overall dynamics. 

\begin{figure}[!htbp]
    \centering
    \includegraphics[width=0.6\textwidth]{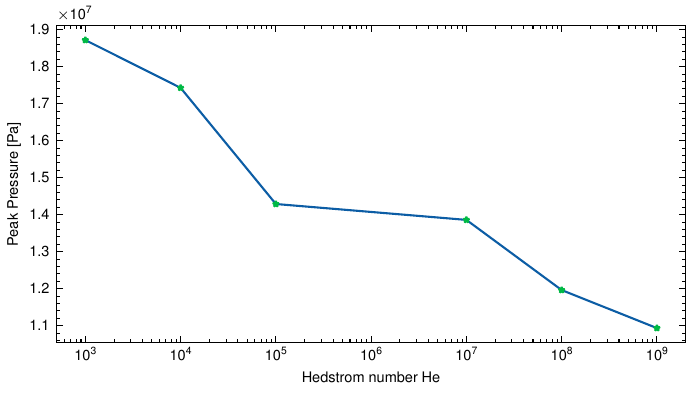} 
    \caption{Maximal over-pressure due to hammer effect against Hedstron number.}
    \label{fig:Peak_He}
\end{figure}

We summarize this finding with Figure \ref{fig:Peak_He}, where the maximal over-pressure for each simulation is shown against the corresponding Hedstrom number. We see how the plot range depicts almost twice overpressure values for the quasi-Newtonian case $\text{He}=10^3$, with respect to $\text{He}=10^9$.

\subsection{Effects of the gradual valve closure}
\label{sec:impact_valve}

\begin{figure}[!htbp]
    \centering
    \includegraphics[width=\textwidth]{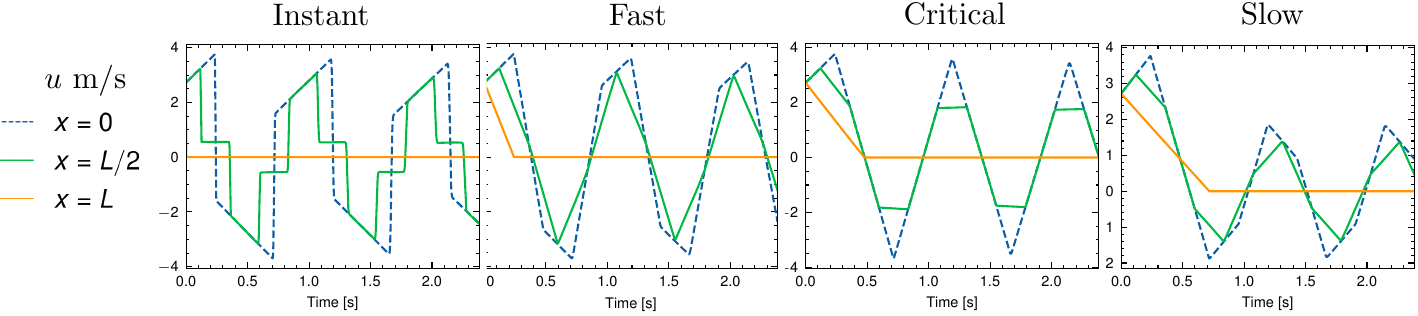}
    \caption{Valve closure time impact on velocity evolution}
    \label{fig:closureModel_vel}
\end{figure}

\begin{figure}[!htbp]
    \centering
    \includegraphics[width=\textwidth]{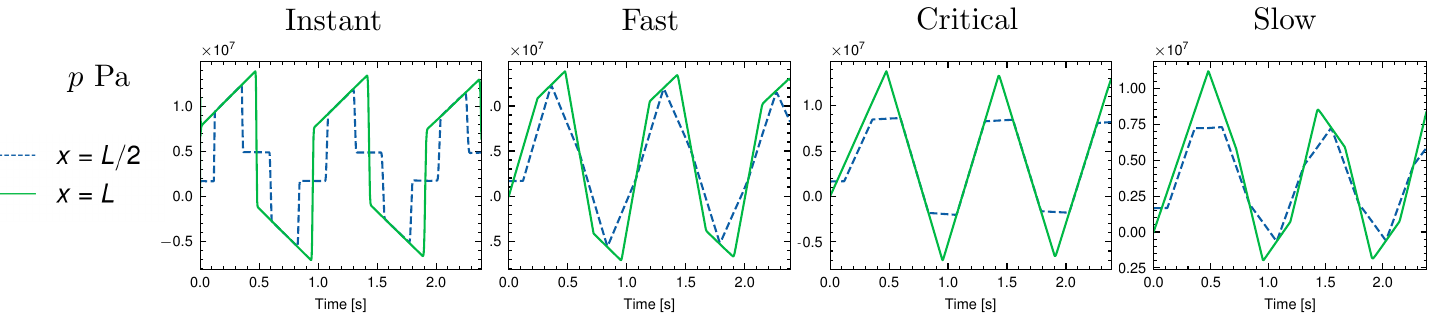}
    \caption{Valve closure time impact on pressure evolution}
    \label{fig:closureModel_press}
\end{figure}

An additional brief test is performed to assess the impact of the gate closure model on the wave dynamics. We set the Dirichlet boundary condition in \eqref{eq:governing_scaled} as:
\begin{equation}
g^*(t^*) = \left(1 - \frac{t^*}{T^*_c}\right),
\end{equation}
where we check several closure times scaling it to \fg{the wave round-trip time $2L/c$. We consider the dimensionless closure time $T^*_c$, meaning that $T^*_c < 2$ will lead to a fast valve closure.} The results, using the parametric configuration for the copper slurry, are shown in Figure \ref{fig:closureModel_vel} for velocity, and Figure \ref{fig:closureModel_press}, for the pressure. We depict four scenarios, namely an instant closure time ($T^*_c = 0$), a fast closure time ($T^*_c = 1$), a critical closure time ($T^*_c = 2$) and a slow closure time ($T^*_c = 3$). As expected, the sharpest numerical solutions of the governing equations are retrieved for the instantaneous closure, and we observe how the delay of the valve operation softens the overall dynamics, still keeping nonetheless similar over-pressure values \fg{in the first peak, yet decreasing at the sub-sequent roundtrips}. Yet, not-sudden pressure changes, as with the case $T^*_c = 3$, are preferable operation conditions.

\begin{figure}[!htbp]
    \centering
    \includegraphics[width=\textwidth]{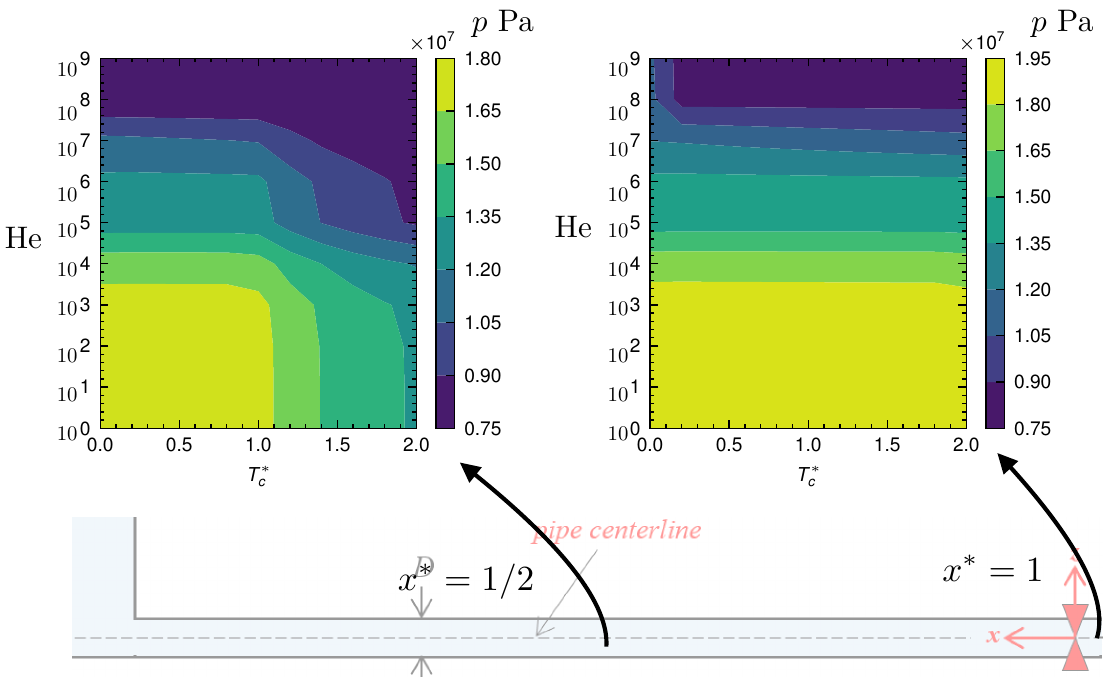}
    \caption{Parametric study for maximal pressure contours against valve closure time and Hedstrom number}
    \label{fig:closureModel_para}
\end{figure}

The \fg{fast} valve closure time study is summarized by looking at the plots of Figure \ref{fig:closureModel_para}, where we see how the overall simulation pressure peak (i.e. the first peak) behaves against both, the Hedstrom number, and the dimensionless closure time. We first confirm that the Newtonian case brings higher pressure peaks, and we also see that the \fg{critical closure reduces considerably the wave peak at the pipe half length control point ($x^*=1/2$)}. That is to say, at the valve closure control point, the peak pressure behavior is almost independent of the closure time \fg{in the fast closure setup}, whereas at the middle control point, we observe a clear separation at the critical time $T_c = 1$, after which the pressure peak decreases almost linearly.

\subsection{\fg{Sensitivity to non-dimensional scaling and convective effects}}
\label{sec:adimensional}

\begin{figure}
    \centering
    \subfigure[Dimensionless velocity evolution over dimensionless time]{
    \includegraphics[width=\linewidth]{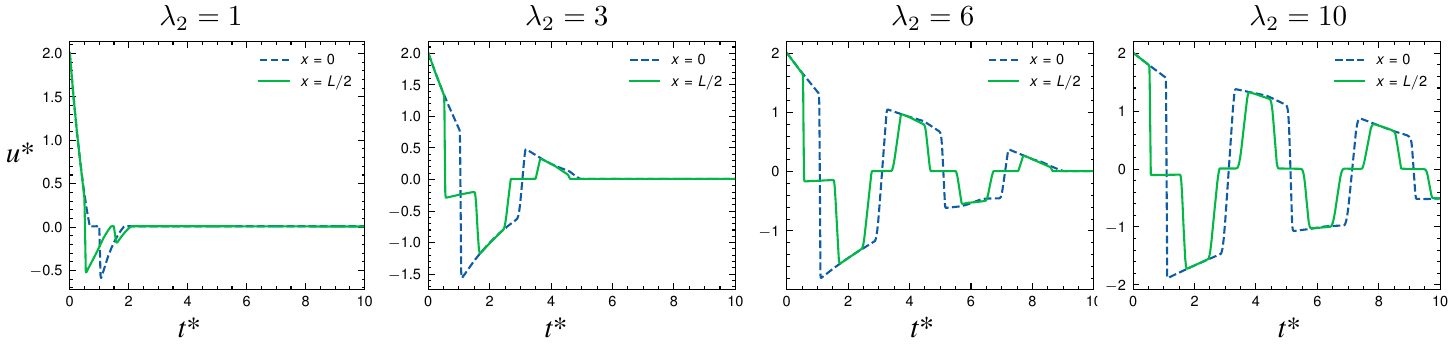}}
    \subfigure[Dimensionless pressure evolution over dimensionless time]{
    \includegraphics[width=\linewidth]{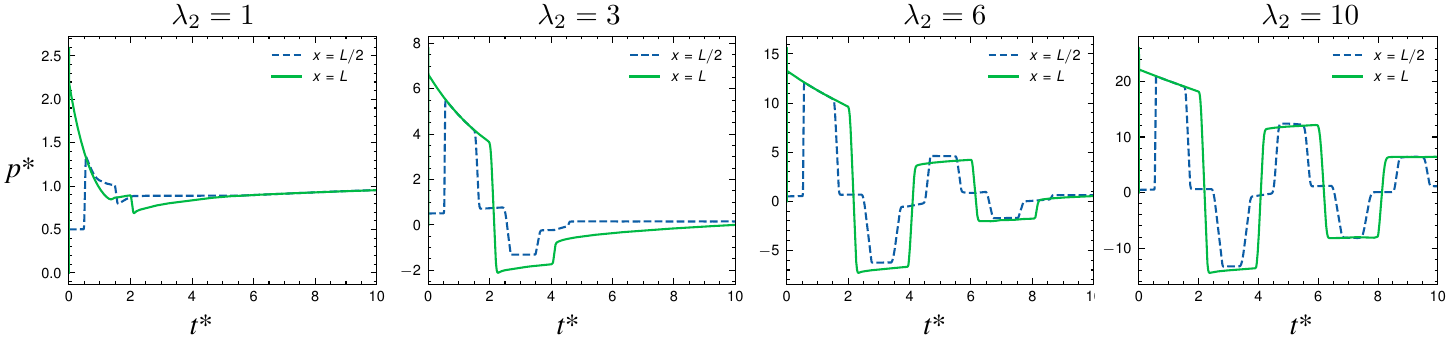}    }
    \caption{\fg{Temporal evolution of the dimensionless solutions of the transient. We move the parameter $\lambda_1$ from 1, where the pressure rise is comparable with the pipe operational conditions (leading to a failed pipe hammer), to 10, where the first pressure peak}}
    \label{fig:lambda_2}
\end{figure}

\fg{We perform an additional study without assuming a specific application set-up. For this experiment, we employ $\widetilde{\Rey} = 5000$ and $\He=10^5$. In Figure \ref{fig:lambda_2} we observe the behavior of the dimensionless solutions with respect to the parameter $\lambda_2$ in Eqs.~\eqref{eq:governing_scaled}, keeping $\lambda_1 = (2/10) \lambda_2$ (thus fixing the Mach number). One should notice that $\lambda_2 = \delta/32 \widetilde{\Rey} \widetilde{Ma}^{-1} = c \rho_m U_0 / P_0$, that is to say, it gives us an idea of how much the pipe hammer over-pressure (calculated with the surrogate Joukowsky formula using the steady Newtonian velocity) exceeds the tank pressure $P_0$.}

\fg{Figure \ref{fig:lambda_2} shows how $\lambda_2 = 1$ leads to an over-pressure which is not significantly different from the operational pressure induced by the tank, yielding a non-oscillatory regime, where the hammer wave is likely irrelevant. On the contrary, the case $\lambda_2=10$ leads to a higher pressure rise, in which the frictional forces are not able to dissipate the wave as quickly as with lower values. Any larger $\lambda_2$ will lead to slower dissipation rates such that friction losses become negligible (as it would be, e.g., for $\lambda_2 = 100$). This is an expected behavior, in spite of the additional resistance imposed by the plastic behavior of the fluid.}

\begin{figure}
    \centering
    \includegraphics[width=\linewidth]{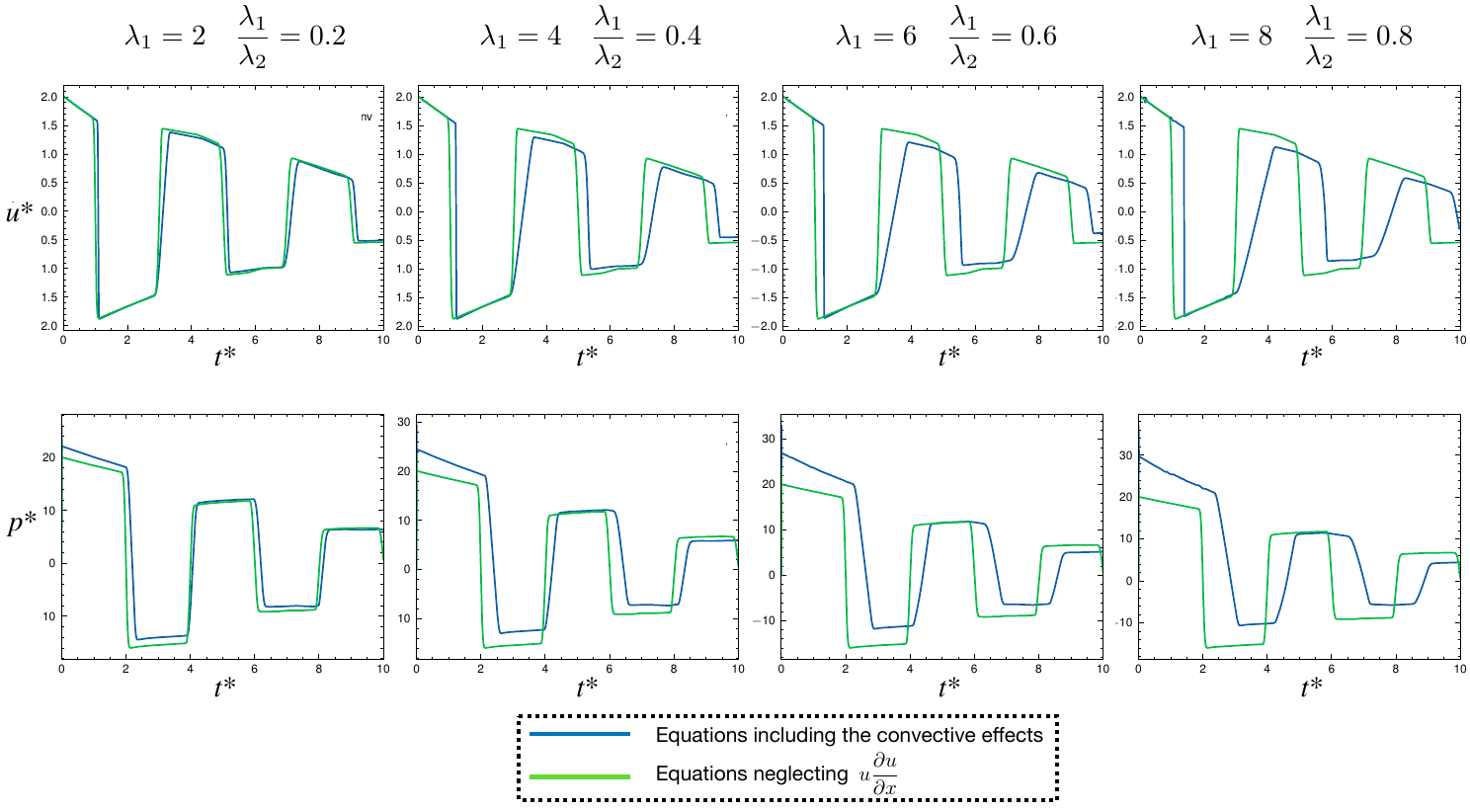}
    \caption{Assessment of the solutions at control points with and without neglecting the convective effects in \eqref{eq:governing_scaled}. We observe how larger differences arise as the ratio $\lambda_1/\lambda_2$ increases.}
    \label{fig:convective}
\end{figure}

\fg{On a different avenue, we fix $\lambda_2 = 10$ and compare simulations for different $\lambda_1$, therefore changing the relative contribution of the convective term $u^* \frac{\partial u^*}{\partial x^*}$ in Eqs.~\eqref{eq:governing_scaled}, and thereby also changing the coefficient $\lambda_2/\lambda_1 = Ma$. We acknowledge that, even though we keep a subsonic regime, our model is unable to reproduce the fully nonlinear equations required by such conditions. This is a step further in our research, where the convective effects are indeed relevant, as shown even in Fig.~\ref{fig:convective} for our surrogate model. These differences between considering or not the convective transport effects, reveal not only a change in the peak pressure of the solutions, but also a temporal shift in their evolution, suggesting larger dissipation times, even at moderately low Mach numbers, as with the cases $\lambda_1=2$ or $\lambda_1=4$. We present this as a step towards  better models for highly non-linear regimes, yet admittedly simplistic for highly compressible scenarios. We emphasize that the methodological proposal of our work (i.e. using FEM) allows this without significant changes in the numerical formulation. This would not be possible with classical methods such as the MOC.}

\section{Conclusions}

\fg{This study provides a parametric analysis of the pipeline hammer phenomenon in non-Newtonian Bingham plastic fluids, employing an efficient computationally inexpensive finite element method (FEM) to accurately simulate the transient dynamics. By integrating the effects of different physical parameters such as yield stress, convective effects, and varying valve closure times, this work advances the understanding of how non-Newtonian rheology affects pressure shockwaves in pipeline systems. The findings underscore some insights:}

\begin{itemize}

\item Impact of Yield Stress on Attenuation: The yield stress in Bingham plastics contributes a damping effect, which facilitates faster attenuation of pressure peaks compared to Newtonian fluids. This effect is especially critical in applications where the reduction of peak pressures is essential for preventing pipeline stress and material fatigue. As the yield stress increases, the number of pressure oscillations declines significantly, leading to enhanced stability within the system.

\item Valve Closure Dynamics: The analysis of valve closure time revealed that gradual valve closures can lead to delayed, slightly attenuated, and smoother pressure waves and peaks. \fg{The parametric analysis is summarized in Figures \ref{fig:closureModel_vel} and \ref{fig:closureModel_press} for instantaneous (ideal), fast, and slow closures. Additionally, we have investigated the variation of peak pressure across configurations in Figure \ref{fig:closureModel_para}, posing our interest in the fast closure scenario for several Bingham fluids.}

\item Adaptive Model for Laminar and Turbulent Flows: The adaptive \fg{quasi-steady} model used for both laminar and turbulent flow regimes \fg{demonstrated that hybrid approaches can be easily incorporated in a finite element formulation.} This adaptability is essential when dealing with fluids that exhibit both Newtonian and non-Newtonian behaviors depending on flow rates and operational conditions, which contrast to regular studies \cite{oliveira2015} where laminar friction models are adopted even in high velocity conditions, as those usually reached in transient phenomena. 

\item \fg{Influence of Convective Effects and Dimensionless Groups: The dimensionless analysis in Section~\ref{sec:adimensional} shows that increasing the parameter $\lambda_1$ enhances the relative importance of convective transport. This leads to higher peak pressures and a noticeable temporal shift in the pressure response. Even in subsonic regimes, these effects are non-negligible, emphasizing the importance of including convective terms when modeling fast transients. This exploratory study motivates the development of more advanced models capable of handling non-linear compressible effects beyond our simplified framework.}
\end{itemize}

Overall, this work highlights the Bingham slurry parametric study, and also the advantages of using advanced numerical techniques, such as FEM, to accurately capture the complex interplay of rheological properties, flow dynamics, and operational changes in pipelines transporting non-Newtonian fluids. \fg{However, our work also presents some limitations important to highlight and that need to be addressed by future studies on the topic. On the one hand, it is necessary to compare these theoretical predictions with real measurements obtained from physical models—or better yet, with transient measurements from large-scale piping systems. Only in this way can we assess the accuracy of the dynamic model proposed in this article. On the other hand, future research could further refine these models through machine learning approaches for rapid parameter optimization, and stabilization techniques could be explored to enhance model robustness. Concerning the governing equations, future research will also encompass the implementation of a fully unsteady friction model for the pipe walls \cite{wahba2013non}. This work lays a foundation for more resilient pipeline designs and informed operational strategies, particularly in industries where non-Newtonian fluids pose unique challenges to system stability and durability.}

\section{Acknowledgments}

The authors acknowledge the School of Civil Engineering for providing the facilities to conduct this research. Felipe Galarce acknowledges the financial support from the funding ANID Fondecyt Regular 1250287, DI Vinci Iniciación PUCV 039.731/2025, and the Cyprus Research and Innovation Foundation: Horizon Europe - 2nd Opportunity OPPTY-MSCA/0125. Francisco Martinez acknowledges the financial support from the funding Puntaje de Corte, DI Investigación PUCV 210.703/2024.


\newpage
\section{Appendix: List of designations and symbols used in this document}

The following list includes the notation for all variables used in the manuscript:\\

\begin{tabular}{@{}ll@{}}
$x$             & streamwise position along the pipeline \\
$t$             & time \\
$u(x,t)$        & mean flow velocity at position $x$, at time $t$ \\
$p(x,t)$        & mean flow pressure at position $x$, at time $t$ \\
$U_o$           & steady-state velocity scale ($= Q/A$) \\
$P_o$           & steady-state pressure scale ($= \rho_m g H$) \\
$c$             & shockwave celerity \\
$g$             & acceleration of gravity \\
$C_V$           & mixture volume concentration \\
$\rho_f$        & fluid density \\
$\rho_s$        & solids density \\
$\rho_m$        & mixture density \\
$\gamma$        & mixture specific gravity ($= \rho_m g$) \\
$\tau_y$        & mixture yield stress \\
$\eta$          & Bingham dynamic viscosity \\
$\Delta P$      & Joukowsky overpressure pulse \\
$H$             & reservoir total head \\
$\Rey$          & Reynolds number \\
$\He$           & Hedstrom number \\
$L$             & pipeline length \\
$D$             & pipeline inner diameter \\
$e$             & pipeline wall thickness \\
$E$             & pipeline elasticity \\
$K_f$           & fluid compressibility modulus \\
$K_s$           & solid compressibility modulus \\
$K_m$           & mixture compressibility modulus \\
$f$             & Fanning friction coefficient \\
$S(\bu)$        & frictional term \\
$g(t)$          & valve closure law \\
$\Delta x$      & spatial step \\
$\Delta t$      & time step \\
$\Delta a^*$    & dimensionless version of generic variable $a$ \\
$P_A$           & maximum allowable pressure of the pipeline \\
$P_v$           & vacuum pressure \\
\end{tabular}

\section*{Statements and Declarations}
The authors declare no competing interest.




\begin{thebibliography}{79}
\ifx \bisbn   \undefined \def \bisbn  #1{ISBN #1}\fi
\ifx \binits  \undefined \def \binits#1{#1}\fi
\ifx \bauthor  \undefined \def \bauthor#1{#1}\fi
\ifx \batitle  \undefined \def \batitle#1{#1}\fi
\ifx \bjtitle  \undefined \def \bjtitle#1{#1}\fi
\ifx \bvolume  \undefined \def \bvolume#1{\textbf{#1}}\fi
\ifx \byear  \undefined \def \byear#1{#1}\fi
\ifx \bissue  \undefined \def \bissue#1{#1}\fi
\ifx \bfpage  \undefined \def \bfpage#1{#1}\fi
\ifx \blpage  \undefined \def \blpage #1{#1}\fi
\ifx \burl  \undefined \def \burl#1{\textsf{#1}}\fi
\ifx \doiurl  \undefined \def \doiurl#1{\url{https://doi.org/#1}}\fi
\ifx \betal  \undefined \def \betal{\textit{et al.}}\fi
\ifx \binstitute  \undefined \def \binstitute#1{#1}\fi
\ifx \binstitutionaled  \undefined \def \binstitutionaled#1{#1}\fi
\ifx \bctitle  \undefined \def \bctitle#1{#1}\fi
\ifx \beditor  \undefined \def \beditor#1{#1}\fi
\ifx \bpublisher  \undefined \def \bpublisher#1{#1}\fi
\ifx \bbtitle  \undefined \def \bbtitle#1{#1}\fi
\ifx \bedition  \undefined \def \bedition#1{#1}\fi
\ifx \bseriesno  \undefined \def \bseriesno#1{#1}\fi
\ifx \blocation  \undefined \def \blocation#1{#1}\fi
\ifx \bsertitle  \undefined \def \bsertitle#1{#1}\fi
\ifx \bsnm \undefined \def \bsnm#1{#1}\fi
\ifx \bsuffix \undefined \def \bsuffix#1{#1}\fi
\ifx \bparticle \undefined \def \bparticle#1{#1}\fi
\ifx \barticle \undefined \def \barticle#1{#1}\fi
\bibcommenthead
\ifx \bconfdate \undefined \def \bconfdate #1{#1}\fi
\ifx \botherref \undefined \def \botherref #1{#1}\fi
\ifx \url \undefined \def \url#1{\textsf{#1}}\fi
\ifx \bchapter \undefined \def \bchapter#1{#1}\fi
\ifx \bbook \undefined \def \bbook#1{#1}\fi
\ifx \bcomment \undefined \def \bcomment#1{#1}\fi
\ifx \oauthor \undefined \def \oauthor#1{#1}\fi
\ifx \citeauthoryear \undefined \def \citeauthoryear#1{#1}\fi
\ifx \endbibitem  \undefined \def \endbibitem {}\fi
\ifx \bconflocation  \undefined \def \bconflocation#1{#1}\fi
\ifx \arxivurl  \undefined \def \arxivurl#1{\textsf{#1}}\fi
\csname PreBibitemsHook\endcsname

\bibitem[\protect\citeauthoryear{Jiang et~al.}{2024}]{Dan_hammer2024}
\begin{barticle}
\bauthor{\bsnm{Jiang}, \binits{D.}},
\bauthor{\bsnm{Zeng}, \binits{C.}},
\bauthor{\bsnm{Lu}, \binits{Q.}},
\bauthor{\bsnm{Guo}, \binits{Q.}}:
\batitle{Water hammer in pipelines based on different friction models}.
\bjtitle{Scientific Reports}
\bvolume{14}(\bissue{1}),
\bfpage{953}
(\byear{2024})
\doiurl{10.1038/s41598-024-51409-9}
\end{barticle}
\endbibitem

\bibitem[\protect\citeauthoryear{Zhang et~al.}{2023}]{Qiaping2023}
\begin{barticle}
\bauthor{\bsnm{Zhang}, \binits{Q.}},
\bauthor{\bsnm{Tian}, \binits{Z.}},
\bauthor{\bsnm{Lu}, \binits{S.}},
\bauthor{\bsnm{Kang}, \binits{H.}}:
\batitle{Numerical simulation of water hammer in pipeline system using
  efficient wave tracking method}.
\bjtitle{Water Resources Management}
\bvolume{37}(\bissue{8}),
\bfpage{3053}--\blpage{3068}
(\byear{2023})
\end{barticle}
\endbibitem

\bibitem[\protect\citeauthoryear{Wylie}{1978}]{wylie1978fluid}
\begin{botherref}
\oauthor{\bsnm{Wylie}, \binits{V.L.} \bsuffix{E.B.and~Streeter}}:
Fluid Transients.
New York, USA
(1978)
\end{botherref}
\endbibitem

\bibitem[\protect\citeauthoryear{Chaudhry}{2014}]{chaudhry2014}
\begin{botherref}
\oauthor{\bsnm{Chaudhry}, \binits{M.H.}}:
Applied Hydraulic Transients.
Springer
(2014).
\doiurl{10.1007/978-1-4614-8538-4}
\end{botherref}
\endbibitem

\bibitem[\protect\citeauthoryear{Ghidaoui}{2004}]{ghidaoui2004}
\begin{barticle}
\bauthor{\bsnm{Ghidaoui}, \binits{M.S.}}:
\batitle{On the fundamental equations of water hammer}.
\bjtitle{Urban Water Journal}
\bvolume{1}(\bissue{2}),
\bfpage{71}--\blpage{83}
(\byear{2004})
\doiurl{10.1080/15730620412331290001}
\end{barticle}
\endbibitem

\bibitem[\protect\citeauthoryear{Ghidaoui et~al.}{2005}]{ghidaoui2005}
\begin{barticle}
\bauthor{\bsnm{Ghidaoui}, \binits{M.S.}},
\bauthor{\bsnm{Zhao}, \binits{M.}},
\bauthor{\bsnm{McInnis}, \binits{D.A.}},
\bauthor{\bsnm{Axworthy}, \binits{D.H.}}:
\batitle{A review of water hammer theory and practice}.
\bjtitle{Appl. Mech. Rev.}
\bvolume{58}(\bissue{1}),
\bfpage{49}--\blpage{76}
(\byear{2005})
\doiurl{10.1115/1.1828050}
\end{barticle}
\endbibitem

\bibitem[\protect\citeauthoryear{Kodura et~al.}{2018}]{kodura2018}
\begin{bchapter}
\bauthor{\bsnm{Kodura}, \binits{A.}},
\bauthor{\bsnm{Kubrak}, \binits{M.}},
\bauthor{\bsnm{Stefanek}, \binits{P.}},
\bauthor{\bsnm{Weinerowska-Bords}, \binits{K.}}:
\bctitle{An experimental investigation of pressure wave celerity during the
  transient slurry flow}.
In: \bbtitle{Free Surface Flows and Transport Processes: 36th International
  School of Hydraulics},
pp. \bfpage{259}--\blpage{269}
(\byear{2018}).
\bcomment{Springer}
\end{bchapter}
\endbibitem

\bibitem[\protect\citeauthoryear{Kodura et~al.}{2019}]{kodura2019}
\begin{barticle}
\bauthor{\bsnm{Kodura}, \binits{A.}},
\bauthor{\bsnm{Weinerowska-Bords}, \binits{K.}},
\bauthor{\bsnm{Artichowicz}, \binits{W.}},
\bauthor{\bsnm{Kubrak}, \binits{M.}},
\bauthor{\bsnm{Stefanek}, \binits{P.}}:
\batitle{In situ verification of numerical model of water hammer in slurries}.
\bjtitle{Journal of Fluids Engineering}
\bvolume{141}(\bissue{8}),
\bfpage{081115}
(\byear{2019})
\doiurl{10.1115/1.4042959}
\end{barticle}
\endbibitem

\bibitem[\protect\citeauthoryear{Kodura et~al.}{2022}]{kodura2022}
\begin{barticle}
\bauthor{\bsnm{Kodura}, \binits{A.}},
\bauthor{\bsnm{Weinerowska-Bords}, \binits{K.}},
\bauthor{\bsnm{Kubrak}, \binits{M.}}:
\batitle{Simplified numerical model for transient flow of slurries at low
  concentration}.
\bjtitle{Energies}
\bvolume{15}(\bissue{19}),
\bfpage{7175}
(\byear{2022})
\doiurl{10.3390/en15197175}
\end{barticle}
\endbibitem

\bibitem[\protect\citeauthoryear{Rocha et~al.}{2008}]{rocha2008}
\begin{botherref}
\oauthor{\bsnm{Rocha}, \binits{L.}},
\oauthor{\bsnm{Oliveira}, \binits{G.}},
\oauthor{\bsnm{Negrao}, \binits{C.O.}},
\oauthor{\bsnm{Franco}, \binits{A.}},
\oauthor{\bsnm{Martins}, \binits{A.}}:
Modelling the start-up flow of well drilling fluids.
Proceedings of ENCIT
(2008)
\end{botherref}
\endbibitem

\bibitem[\protect\citeauthoryear{Ihle}{2014}]{ihle2014}
\begin{barticle}
\bauthor{\bsnm{Ihle}, \binits{C.F.}}:
\batitle{Should maximum pressures in ore pipelines be computed out of system
  startups or power outages?}
\bjtitle{Minerals Engineering}
\bvolume{55},
\bfpage{57}--\blpage{59}
(\byear{2014})
\doiurl{10.1016/j.mineng.2013.09.006}
\end{barticle}
\endbibitem

\bibitem[\protect\citeauthoryear{Boger}{2013}]{boger2013}
\begin{barticle}
\bauthor{\bsnm{Boger}, \binits{D.V.}}:
\batitle{Rheology of slurries and environmental impacts in the mining
  industry}.
\bjtitle{Annual Review of Chemical and Biomolecular engineering}
\bvolume{4},
\bfpage{239}--\blpage{257}
(\byear{2013})
\doiurl{10.1146/annurev-chembioeng-061312-103347}
\end{barticle}
\endbibitem

\bibitem[\protect\citeauthoryear{Abdul-Wahab and Marikar}{2012}]{abdul2012}
\begin{barticle}
\bauthor{\bsnm{Abdul-Wahab}, \binits{S.}},
\bauthor{\bsnm{Marikar}, \binits{F.}}:
\batitle{The environmental impact of gold mines: pollution by heavy metals}.
\bjtitle{Open Engineering}
\bvolume{2}(\bissue{2}),
\bfpage{304}--\blpage{313}
(\byear{2012})
\doiurl{10.2478/s13531-011-0052-3}
\end{barticle}
\endbibitem

\bibitem[\protect\citeauthoryear{Younger}{2004}]{younger2004}
\begin{barticle}
\bauthor{\bsnm{Younger}, \binits{P.L.}}:
\batitle{Environmental impacts of coal mining and associated wastes: a
  geochemical perspective}.
\bjtitle{Geological Society, London, Special Publications}
\bvolume{236}(\bissue{1}),
\bfpage{169}--\blpage{209}
(\byear{2004})
\doiurl{10.1144/GSL.SP.2004.236.01.12}
\end{barticle}
\endbibitem

\bibitem[\protect\citeauthoryear{Cacciuttolo and Cano}{2022}]{cacciuttolo2022}
\begin{barticle}
\bauthor{\bsnm{Cacciuttolo}, \binits{C.}},
\bauthor{\bsnm{Cano}, \binits{D.}}:
\batitle{Environmental impact assessment of mine tailings spill considering
  metallurgical processes of gold and copper mining: case studies in the
  {A}ndean countries of {C}hile and {P}eru}.
\bjtitle{Water}
\bvolume{14}(\bissue{19}),
\bfpage{3057}
(\byear{2022})
\doiurl{10.3390/w14193057}
\end{barticle}
\endbibitem

\bibitem[\protect\citeauthoryear{Wahba}{2013}]{wahba2013non}
\begin{barticle}
\bauthor{\bsnm{Wahba}, \binits{E.}}:
\batitle{Non-newtonian fluid hammer in elastic circular pipes: Shear-thinning
  and shear-thickening effects}.
\bjtitle{Journal of Non-Newtonian Fluid Mechanics}
\bvolume{198},
\bfpage{24}--\blpage{30}
(\byear{2013})
\doiurl{10.1016/j.jnnfm.2013.04.007}
\end{barticle}
\endbibitem

\bibitem[\protect\citeauthoryear{Baha~Abulnaga}{2021}]{abulnaga2021}
\begin{botherref}
\oauthor{\bsnm{Baha~Abulnaga}, \binits{P.E.}}:
Slurry Systems Handbook.
McGraw-Hill Education
(2021)
\end{botherref}
\endbibitem

\bibitem[\protect\citeauthoryear{O’brien and Julien}{1985}]{obrien1985}
\begin{botherref}
\oauthor{\bsnm{O’brien}, \binits{J.}},
\oauthor{\bsnm{Julien}, \binits{P.}}:
Physical properties and mechanics of hyperconcentrated sediment flows.
Proc. ASCE HD Delineation of landslides, flash flood and debris flow Hazards
(1985)
\end{botherref}
\endbibitem

\bibitem[\protect\citeauthoryear{O'Brien and Julien}{1988}]{obrien1988}
\begin{barticle}
\bauthor{\bsnm{O'Brien}, \binits{J.S.}},
\bauthor{\bsnm{Julien}, \binits{P.Y.}}:
\batitle{Laboratory analysis of mudflow properties}.
\bjtitle{Journal of hydraulic engineering}
\bvolume{114}(\bissue{8}),
\bfpage{877}--\blpage{887}
(\byear{1988})
\doiurl{10.1061/(ASCE)0733-9429(1988)114:8(877)}
\end{barticle}
\endbibitem

\bibitem[\protect\citeauthoryear{Julien and Lan}{1991}]{julien1991}
\begin{barticle}
\bauthor{\bsnm{Julien}, \binits{P.Y.}},
\bauthor{\bsnm{Lan}, \binits{Y.}}:
\batitle{Rheology of hyperconcentrations}.
\bjtitle{Journal of Hydraulic Engineering}
\bvolume{117}(\bissue{3}),
\bfpage{346}--\blpage{353}
(\byear{1991})
\doiurl{10.1061/(ASCE)0733-9429(1991)117:3(346)}
\end{barticle}
\endbibitem

\bibitem[\protect\citeauthoryear{Wasp et~al.}{1977}]{wasp1977}
\begin{botherref}
\oauthor{\bsnm{Wasp}, \binits{E.J.}},
\oauthor{\bsnm{Kenny}, \binits{J.P.}},
\oauthor{\bsnm{Gandhi}, \binits{R.L.}}:
Solid-liquid flow: Slurry pipeline transportation [pumps, valves, mechanical
  equipment, economics].
Ser. Bulk Mater. Handl.;(United States)
\textbf{1}(4)
(1977)
\end{botherref}
\endbibitem

\bibitem[\protect\citeauthoryear{Mitishita et~al.}{2018}]{mitishita2018}
\begin{barticle}
\bauthor{\bsnm{Mitishita}, \binits{R.S.}},
\bauthor{\bsnm{Oliveira}, \binits{G.M.}},
\bauthor{\bsnm{Santos}, \binits{T.G.}},
\bauthor{\bsnm{Negr{\~a}o}, \binits{C.O.}}:
\batitle{Pressure transmission in yield stress fluids-an experimental
  analysis}.
\bjtitle{Journal of Non-Newtonian Fluid Mechanics}
\bvolume{261},
\bfpage{50}--\blpage{59}
(\byear{2018})
\end{barticle}
\endbibitem

\bibitem[\protect\citeauthoryear{Pham and Choi}{2021}]{PHAM2021121099}
\begin{barticle}
\bauthor{\bsnm{Pham}, \binits{T.Q.D.}},
\bauthor{\bsnm{Choi}, \binits{S.}}:
\batitle{Numerical analysis of direct contact condensation-induced water
  hammering effect using openfoam in realistic steam pipes}.
\bjtitle{International Journal of Heat and Mass Transfer}
\bvolume{171},
\bfpage{121099}
(\byear{2021})
\doiurl{10.1016/j.ijheatmasstransfer.2021.121099}
\end{barticle}
\endbibitem

\bibitem[\protect\citeauthoryear{Darbhamulla and
  Jaiman}{2024}]{DARBHAMULLA2024106283}
\begin{barticle}
\bauthor{\bsnm{Darbhamulla}, \binits{N.B.}},
\bauthor{\bsnm{Jaiman}, \binits{R.K.}}:
\batitle{A finite element framework for fluid–structure interaction of
  turbulent cavitating flows with flexible structures}.
\bjtitle{Computers \& Fluids}
\bvolume{277},
\bfpage{106283}
(\byear{2024})
\doiurl{10.1016/j.compfluid.2024.106283}
\end{barticle}
\endbibitem

\bibitem[\protect\citeauthoryear{Wang et~al.}{2017}]{WANG20171}
\begin{barticle}
\bauthor{\bsnm{Wang}, \binits{C.}},
\bauthor{\bsnm{Nilsson}, \binits{H.}},
\bauthor{\bsnm{Yang}, \binits{J.}},
\bauthor{\bsnm{Petit}, \binits{O.}}:
\batitle{1d–3d coupling for hydraulic system transient simulations}.
\bjtitle{Computer Physics Communications}
\bvolume{210},
\bfpage{1}--\blpage{9}
(\byear{2017})
\doiurl{10.1016/j.cpc.2016.09.007}
\end{barticle}
\endbibitem

\bibitem[\protect\citeauthoryear{Oliveira et~al.}{2015}]{oliveira2015}
\begin{barticle}
\bauthor{\bsnm{Oliveira}, \binits{G.M.}},
\bauthor{\bsnm{Franco}, \binits{A.T.}},
\bauthor{\bsnm{Negrão}, \binits{C.O.R.}}:
\batitle{{Mathematical Model for Viscoplastic Fluid Hammer}}.
\bjtitle{Journal of Fluids Engineering}
\bvolume{138}(\bissue{1}),
\bfpage{011301}
(\byear{2015})
\doiurl{10.1115/1.4031001}
{\href{https://arxiv.org/abs/https://asmedigitalcollection.asme.org/fluidsengineering/article-pdf/138/1/011301/6194304/fe\_138\_01\_011301.pdf}{{https://asmedigitalcollection.asme.org/fluidsengineering/article-pdf/138/1/011301/6194304/fe\_138\_01\_011301.pdf}}}
\end{barticle}
\endbibitem

\bibitem[\protect\citeauthoryear{Tawfik}{2023}]{hammerIA}
\begin{barticle}
\bauthor{\bsnm{Tawfik}, \binits{A.}}:
\batitle{Air vessel sizing approach for pipeline protection using artificial
  neural networks}.
\bjtitle{Journal of Engineering and Applied Science}
\bvolume{70}(\bissue{1}),
\bfpage{34}
(\byear{2023})
\doiurl{10.1186/s44147-023-00206-8}
\end{barticle}
\endbibitem

\bibitem[\protect\citeauthoryear{Brunton and Kutz}{2019}]{Brunton_Kutz_2019}
\begin{botherref}
\oauthor{\bsnm{Brunton}, \binits{S.L.}},
\oauthor{\bsnm{Kutz}, \binits{J.N.}}:
Data-Driven Science and Engineering: Machine Learning, Dynamical Systems, and
  Control.
Cambridge University Press
(2019)
\end{botherref}
\endbibitem

\bibitem[\protect\citeauthoryear{Fortier}{1967}]{fortier1967}
\begin{botherref}
\oauthor{\bsnm{Fortier}, \binits{A.}}:
M{\'e}canique des Suspensions.
Masson
(1967)
\end{botherref}
\endbibitem

\bibitem[\protect\citeauthoryear{Adams et~al.}{2023}]{adams2023}
\begin{barticle}
\bauthor{\bsnm{Adams}, \binits{I.}},
\bauthor{\bsnm{Simeonov}, \binits{J.}},
\bauthor{\bsnm{Bateman}, \binits{S.}},
\bauthor{\bsnm{Keane}, \binits{N.}}:
\batitle{A bingham plastic fluid solver for turbulent flow of dense muddy
  sediment mixtures}.
\bjtitle{Fluids}
\bvolume{8}(\bissue{6}),
\bfpage{171}
(\byear{2023})
\doiurl{10.3390/fluids8060171}
\end{barticle}
\endbibitem

\bibitem[\protect\citeauthoryear{De~Oliveira et~al.}{2010}]{oliveira2010}
\begin{barticle}
\bauthor{\bsnm{De~Oliveira}, \binits{G.M.}},
\bauthor{\bsnm{Rocha}, \binits{L.L.V.}},
\bauthor{\bsnm{Franco}, \binits{A.T.}},
\bauthor{\bsnm{Negr{\~a}o}, \binits{C.O.}}:
\batitle{Numerical simulation of the start-up of bingham fluid flows in
  pipelines}.
\bjtitle{Journal of Non-Newtonian Fluid Mechanics}
\bvolume{165}(\bissue{19-20}),
\bfpage{1114}--\blpage{1128}
(\byear{2010})
\doiurl{10.1016/j.jnnfm.2010.05.009}
\end{barticle}
\endbibitem

\bibitem[\protect\citeauthoryear{Oliveira et~al.}{2012}]{oliveira2012}
\begin{barticle}
\bauthor{\bsnm{Oliveira}, \binits{G.M.}},
\bauthor{\bsnm{Negr{\~a}o}, \binits{C.O.}},
\bauthor{\bsnm{Franco}, \binits{A.T.}}:
\batitle{Pressure transmission in bingham fluids compressed within a closed
  pipe}.
\bjtitle{Journal of Non-Newtonian Fluid Mechanics}
\bvolume{169},
\bfpage{121}--\blpage{125}
(\byear{2012})
\doiurl{10.1016/j.jnnfm.2018.08.007}
\end{barticle}
\endbibitem

\bibitem[\protect\citeauthoryear{Santos et~al.}{2023}]{santos2023}
\begin{barticle}
\bauthor{\bsnm{Santos}, \binits{T.G.}},
\bauthor{\bsnm{Oliveira}, \binits{G.M.}},
\bauthor{\bsnm{Negrao}, \binits{C.O.}}:
\batitle{Dimensionless analysis of non-{N}ewtonian power-law fluid hammer}.
\bjtitle{Journal of Hydraulic Engineering}
\bvolume{149}(\bissue{9}),
\bfpage{04023034}
(\byear{2023})
\doiurl{10.1061/JHEND8.HYENG-13276}
\end{barticle}
\endbibitem

\bibitem[\protect\citeauthoryear{Abugattas et~al.}{2020}]{Abugattas2020}
\begin{barticle}
\bauthor{\bsnm{Abugattas}, \binits{C.}},
\bauthor{\bsnm{Aguirre}, \binits{A.}},
\bauthor{\bsnm{Castillo}, \binits{E.}},
\bauthor{\bsnm{Cruchaga}, \binits{M.}}:
\batitle{{Numerical study of bifurcation blood flows using three different
  non-{N}ewtonian constitutive models}}.
\bjtitle{Applied Mathematical Modelling}
\bvolume{88},
\bfpage{529}--\blpage{549}
(\byear{2020})
\doiurl{10.1016/j.apm.2020.06.066}
\end{barticle}
\endbibitem

\bibitem[\protect\citeauthoryear{{Monteiro Andrade}
  et~al.}{2023}]{MONTEIROANDRADE2023104391}
\begin{barticle}
\bauthor{\bsnm{{Monteiro Andrade}}, \binits{D.}},
\bauthor{\bsnm{{Bastos de Freitas Rachid}}, \binits{F.}},
\bauthor{\bsnm{Sieno~Tijsseling}, \binits{A.}}:
\batitle{A thermodynamically consistent model for hydraulic transients in
  metallic pipes undergoing elasto-viscoplastic deformations}.
\bjtitle{International Journal of Non-Linear Mechanics}
\bvolume{152},
\bfpage{104391}
(\byear{2023})
\doiurl{10.1016/j.ijnonlinmec.2023.104391}
\end{barticle}
\endbibitem

\bibitem[\protect\citeauthoryear{White}{1991}]{white1991}
\begin{botherref}
\oauthor{\bsnm{White}, \binits{F.M.}}:
Viscous Fluid Flow.
McGraw-Hill
(1991)
\end{botherref}
\endbibitem

\bibitem[\protect\citeauthoryear{Duan et~al.}{2020}]{duan2020_waterUWSS}
\begin{barticle}
\bauthor{\bsnm{Duan}, \binits{H.-F.}},
\bauthor{\bsnm{Pan}, \binits{B.}},
\bauthor{\bsnm{Wang}, \binits{M.}},
\bauthor{\bsnm{Chen}, \binits{L.}},
\bauthor{\bsnm{Zheng}, \binits{F.}},
\bauthor{\bsnm{Zhang}, \binits{Y.}}:
\batitle{State-of-the-art review on the transient flow modeling and utilization
  for urban water supply system (uwss) management}.
\bjtitle{Journal of Water Supply: Research and Technology-Aqua}
\bvolume{69}(\bissue{8}),
\bfpage{858}--\blpage{893}
(\byear{2020})
\doiurl{10.2166/aqua.2020.048}
{\href{https://arxiv.org/abs/https://iwaponline.com/aqua/article-pdf/69/8/858/824143/jws0690858.pdf}{{https://iwaponline.com/aqua/article-pdf/69/8/858/824143/jws0690858.pdf}}}
\end{barticle}
\endbibitem

\bibitem[\protect\citeauthoryear{Plouraboué}{2024}]{PLOURABOUE2024237}
\begin{barticle}
\bauthor{\bsnm{Plouraboué}, \binits{F.}}:
\batitle{Review on water-hammer waves mechanical and theoretical foundations}.
\bjtitle{European Journal of Mechanics - B/Fluids}
\bvolume{108},
\bfpage{237}--\blpage{271}
(\byear{2024})
\doiurl{10.1016/j.euromechflu.2024.08.001}
\end{barticle}
\endbibitem

\bibitem[\protect\citeauthoryear{Riasi et~al.}{2009}]{riasi2009}
\begin{barticle}
\bauthor{\bsnm{Riasi}, \binits{A.}},
\bauthor{\bsnm{Nourbakhsh}, \binits{A.}},
\bauthor{\bsnm{Raisee}, \binits{M.}}:
\batitle{Unsteady velocity profiles in laminar and turbulent water hammer
  flows}.
\bjtitle{Journal of Fluids Engineering}
\bvolume{131}(\bissue{12}),
\bfpage{121202}
(\byear{2009})
\doiurl{10.1115/1.4000557}
{\href{https://arxiv.org/abs/https://asmedigitalcollection.asme.org/fluidsengineering/article-pdf/131/12/121202/5837167/121202\_1.pdf}{{https://asmedigitalcollection.asme.org/fluidsengineering/article-pdf/131/12/121202/5837167/121202\_1.pdf}}}
\end{barticle}
\endbibitem

\bibitem[\protect\citeauthoryear{Urbanowicz
  et~al.}{2023}]{URBANOWICZ2023117848}
\begin{barticle}
\bauthor{\bsnm{Urbanowicz}, \binits{K.}},
\bauthor{\bsnm{Bergant}, \binits{A.}},
\bauthor{\bsnm{Stosiak}, \binits{M.}},
\bauthor{\bsnm{Karpenko}, \binits{M.}},
\bauthor{\bsnm{Bogdevičius}, \binits{M.}}:
\batitle{Developments in analytical wall shear stress modelling for water
  hammer phenomena}.
\bjtitle{Journal of Sound and Vibration}
\bvolume{562},
\bfpage{117848}
(\byear{2023})
\doiurl{10.1016/j.jsv.2023.117848}
\end{barticle}
\endbibitem

\bibitem[\protect\citeauthoryear{Gonzaga~Filho and Bastos~de
  Freitas~Rachid}{2023}]{filho2023}
\begin{botherref}
\oauthor{\bsnm{Gonzaga~Filho}, \binits{J.}},
\oauthor{\bsnm{Freitas~Rachid}, \binits{F.}}:
Comparative analysis of unsteady friction models for pipe flows in light of the
  second law of thermodynamics.
Journal of the Brazilian Society of Mechanical Sciences and Engineering
\textbf{45}
(2023)
\doiurl{10.1007/s40430-022-04007-7}
\end{botherref}
\endbibitem

\bibitem[\protect\citeauthoryear{Urbanowicz et~al.}{2022}]{urbanoWater}
\begin{botherref}
\oauthor{\bsnm{Urbanowicz}, \binits{K.}},
\oauthor{\bsnm{Bergant}, \binits{A.}},
\oauthor{\bsnm{Stosiak}, \binits{M.}},
\oauthor{\bsnm{Deptuła}, \binits{A.}},
\oauthor{\bsnm{Karpenko}, \binits{M.}},
\oauthor{\bsnm{Kubrak}, \binits{M.}},
\oauthor{\bsnm{Kodura}, \binits{A.}}:
Water hammer simulation using simplified convolution-based unsteady friction
  model.
Water
\textbf{14}
(2022)
\doiurl{10.3390/w14193151}
\end{botherref}
\endbibitem

\bibitem[\protect\citeauthoryear{Xu et~al.}{2025}]{XU2025123779}
\begin{barticle}
\bauthor{\bsnm{Xu}, \binits{Y.}},
\bauthor{\bsnm{Zhou}, \binits{L.}},
\bauthor{\bsnm{Lu}, \binits{Y.}},
\bauthor{\bsnm{Hu}, \binits{Y.}},
\bauthor{\bsnm{Zhang}, \binits{Y.}}:
\batitle{Physics-informed neural networks involving unsteady friction for
  transient pipe flow}.
\bjtitle{Water Research}
\bvolume{283},
\bfpage{123779}
(\byear{2025})
\doiurl{10.1016/j.watres.2025.123779}
\end{barticle}
\endbibitem

\bibitem[\protect\citeauthoryear{Raissi et~al.}{2019}]{RAISSI2019686}
\begin{barticle}
\bauthor{\bsnm{Raissi}, \binits{M.}},
\bauthor{\bsnm{Perdikaris}, \binits{P.}},
\bauthor{\bsnm{Karniadakis}, \binits{G.E.}}:
\batitle{Physics-informed neural networks: A deep learning framework for
  solving forward and inverse problems involving nonlinear partial differential
  equations}.
\bjtitle{Journal of Computational Physics}
\bvolume{378},
\bfpage{686}--\blpage{707}
(\byear{2019})
\doiurl{10.1016/j.jcp.2018.10.045}
\end{barticle}
\endbibitem

\bibitem[\protect\citeauthoryear{Maday et~al.}{2015}]{MPPY2015}
\begin{barticle}
\bauthor{\bsnm{Maday}, \binits{Y.}},
\bauthor{\bsnm{Patera}, \binits{A.T.}},
\bauthor{\bsnm{Penn}, \binits{J.D.}},
\bauthor{\bsnm{Yano}, \binits{M.}}:
\batitle{A parameterized-background data-weak approach to variational data
  assimilation: formulation, analysis, and application to acoustics}.
\bjtitle{International Journal for Numerical Methods in Engineering}
\bvolume{102}(\bissue{5}),
\bfpage{933}--\blpage{965}
(\byear{2015})
\doiurl{10.1002/nme.4747}
{\href{https://arxiv.org/abs/https://onlinelibrary.wiley.com/doi/pdf/10.1002/nme.4747}{{https://onlinelibrary.wiley.com/doi/pdf/10.1002/nme.4747}}}
\end{barticle}
\endbibitem

\bibitem[\protect\citeauthoryear{Galarce et~al.}{2025}]{galarce2023bias}
\begin{barticle}
\bauthor{\bsnm{Galarce}, \binits{F.}},
\bauthor{\bsnm{Mura}, \binits{J.}},
\bauthor{\bsnm{Caiazzo}, \binits{A.}}:
\batitle{Bias and multiscale correction methods for variational state
  estimation}.
\bjtitle{Applied Mathematical Modelling}
\bvolume{138},
\bfpage{115761}
(\byear{2025})
\doiurl{10.1016/j.apm.2024.115761}
\end{barticle}
\endbibitem

\bibitem[\protect\citeauthoryear{Urbanowicz et~al.}{2023}]{urbanowicz_2023}
\begin{barticle}
\bauthor{\bsnm{Urbanowicz}, \binits{K.}},
\bauthor{\bsnm{Haluch}, \binits{I.}},
\bauthor{\bsnm{Bergant}, \binits{A.}},
\bauthor{\bsnm{Deptuła}, \binits{A.}},
\bauthor{\bsnm{Śliwiński}, \binits{P.}}:
\batitle{Initial investigation of wave interactions during simultaneous valve
  closures in hydraulic piping systems}.
\bjtitle{Water Resources Management}
\bvolume{37},
\bfpage{5105}--\blpage{5125}
(\byear{2023})
\doiurl{10.1007/s11269-023-03597-8}
\end{barticle}
\endbibitem

\bibitem[\protect\citeauthoryear{Elkhatib et~al.}{2022}]{ELKHATIB2022139}
\begin{barticle}
\bauthor{\bsnm{Elkhatib}, \binits{H.}},
\bauthor{\bsnm{Alyan}, \binits{A.}},
\bauthor{\bsnm{Abdelrahman}, \binits{A.}}:
\batitle{Hydraulic transient simulation of primary cooling circuit in egyptian
  second research reactor}.
\bjtitle{Journal of King Saud University - Engineering Sciences}
\bvolume{34}(\bissue{2}),
\bfpage{139}--\blpage{146}
(\byear{2022})
\doiurl{10.1016/j.jksues.2020.09.001}
\end{barticle}
\endbibitem

\bibitem[\protect\citeauthoryear{Swamee and Aggarwal}{2011}]{SWAMEE2011178}
\begin{barticle}
\bauthor{\bsnm{Swamee}, \binits{P.K.}},
\bauthor{\bsnm{Aggarwal}, \binits{N.}}:
\batitle{Explicit equations for laminar flow of bingham plastic fluids}.
\bjtitle{Journal of Petroleum Science and Engineering}
\bvolume{76}(\bissue{3}),
\bfpage{178}--\blpage{184}
(\byear{2011})
\doiurl{10.1016/j.petrol.2011.01.015}
\end{barticle}
\endbibitem

\bibitem[\protect\citeauthoryear{Darby and Melson}{1981}]{darby1981}
\begin{botherref}
\oauthor{\bsnm{Darby}, \binits{R.}},
\oauthor{\bsnm{Melson}, \binits{J.}}:
How to predict the friction factor for flow of bingham plastics.
Chemical Engineering
(1981)
\end{botherref}
\endbibitem

\bibitem[\protect\citeauthoryear{Darby et~al.}{1992}]{darby1992}
\begin{barticle}
\bauthor{\bsnm{Darby}, \binits{R.}},
\bauthor{\bsnm{Mun}, \binits{R.}},
\bauthor{\bsnm{Boger}, \binits{D.}}:
\batitle{Predict friction loss in slurry pipes}.
\bjtitle{Chemical engineering}
\bvolume{99}(\bissue{9}),
\bfpage{116}--\blpage{119}
(\byear{1992})
\end{barticle}
\endbibitem

\bibitem[\protect\citeauthoryear{Tijsseling and
  Anderson}{2007}]{tijsseling2007}
\begin{barticle}
\bauthor{\bsnm{Tijsseling}, \binits{A.S.}},
\bauthor{\bsnm{Anderson}, \binits{A.}}:
\batitle{Johannes von kries and the history of water hammer}.
\bjtitle{Journal of Hydraulic Engineering}
\bvolume{133}(\bissue{1}),
\bfpage{1}--\blpage{8}
(\byear{2007})
\end{barticle}
\endbibitem

\bibitem[\protect\citeauthoryear{Young}{1808}]{young1808}
\begin{botherref}
\oauthor{\bsnm{Young}, \binits{T.}}:
Xiii. hydraulic investigations, subservient to an intended croonian lecture on
  the motion of the blood.
Philosophical Transactions of the Royal society of London
(98),
164--186
(1808)
\end{botherref}
\endbibitem

\bibitem[\protect\citeauthoryear{Korteweg}{1878a}]{korteweg1878a}
\begin{botherref}
\oauthor{\bsnm{Korteweg}, \binits{D.J.}}:
Over voortplantings-snelheid van golven in elastische buizen.
Van Doesburgh, Netherlands
(1878)
\end{botherref}
\endbibitem

\bibitem[\protect\citeauthoryear{Korteweg}{1878b}]{korteweg1878b}
\begin{barticle}
\bauthor{\bsnm{Korteweg}, \binits{D.J.}}:
\batitle{Ueber die fortpflanzungsgeschwindigkeit des schalles in elastischen
  r{\"o}hren}.
\bjtitle{Annalen der Physik}
\bvolume{241}(\bissue{12}),
\bfpage{525}--\blpage{542}
(\byear{1878})
\end{barticle}
\endbibitem

\bibitem[\protect\citeauthoryear{Wylie et~al.}{1993}]{streeter1993}
\begin{botherref}
\oauthor{\bsnm{Wylie}, \binits{E.B.}},
\oauthor{\bsnm{Streeter}, \binits{V.L.}},
\oauthor{\bsnm{Suo}, \binits{L.}}:
Fluid Transients in Systems.
Pearson College Div
(1993)
\end{botherref}
\endbibitem

\bibitem[\protect\citeauthoryear{Thorley and Hwang}{1979}]{thorley1979}
\begin{bchapter}
\bauthor{\bsnm{Thorley}, \binits{A.}},
\bauthor{\bsnm{Hwang}, \binits{L.}}:
\bctitle{Effects of rapid change in flowrate of solid-liquid mixtures}.
In: \bbtitle{Proceedings of Hydrotransport 6th Conference, UK},
pp. \bfpage{229}--\blpage{242}
(\byear{1979})
\end{bchapter}
\endbibitem

\bibitem[\protect\citeauthoryear{Han et~al.}{1998}]{han1998}
\begin{barticle}
\bauthor{\bsnm{Han}, \binits{W.}},
\bauthor{\bsnm{Dong}, \binits{Z.}},
\bauthor{\bsnm{Chai}, \binits{H.}}:
\batitle{Water hammer in pipelines with hyperconcentrated slurry flows carrying
  solid particles}.
\bjtitle{Science in China Series E: Technological Sciences}
\bvolume{41},
\bfpage{337}--\blpage{347}
(\byear{1998})
\end{barticle}
\endbibitem

\bibitem[\protect\citeauthoryear{Hanks}{1967}]{hanks1967}
\begin{barticle}
\bauthor{\bsnm{Hanks}, \binits{R.W.}}:
\batitle{On the flow of bingham plastic slurries in pipes and between parallel
  plates}.
\bjtitle{Society of Petroleum Engineers Journal}
\bvolume{7}(\bissue{04}),
\bfpage{342}--\blpage{346}
(\byear{1967})
\end{barticle}
\endbibitem

\bibitem[\protect\citeauthoryear{Assefa and Kaushal}{2015}]{assefa2015}
\begin{botherref}
\oauthor{\bsnm{Assefa}, \binits{K.M.}},
\oauthor{\bsnm{Kaushal}, \binits{D.R.}}:
A comparative study of friction factor correlations for high concentrate slurry
  flow in smooth pipes.
Journal of Hydrology and Hydromechanics
\textbf{63}
(2015)
\end{botherref}
\endbibitem

\bibitem[\protect\citeauthoryear{Kerger et~al.}{2011}]{KERGER20112030}
\begin{barticle}
\bauthor{\bsnm{Kerger}, \binits{F.}},
\bauthor{\bsnm{Archambeau}, \binits{P.}},
\bauthor{\bsnm{Erpicum}, \binits{S.}},
\bauthor{\bsnm{Dewals}, \binits{B.J.}},
\bauthor{\bsnm{Pirotton}, \binits{M.}}:
\batitle{An exact riemann solver and a godunov scheme for simulating highly
  transient mixed flows}.
\bjtitle{Journal of Computational and Applied Mathematics}
\bvolume{235}(\bissue{8}),
\bfpage{2030}--\blpage{2040}
(\byear{2011})
\doiurl{10.1016/j.cam.2010.09.026}
\end{barticle}
\endbibitem

\bibitem[\protect\citeauthoryear{Zienkiewicz et~al.}{2013}]{fem_bible}
\begin{botherref}
\oauthor{\bsnm{Zienkiewicz}, \binits{O.C.}},
\oauthor{\bsnm{Taylor}, \binits{R.L.}},
\oauthor{\bsnm{Zhu}, \binits{J.Z.}}:
The Finite Element Method: Its Basis and Fundamentals.
ElSevier
(2013)
\end{botherref}
\endbibitem

\bibitem[\protect\citeauthoryear{Ern and Guermond}{2021}]{fem_ern}
\begin{botherref}
\oauthor{\bsnm{Ern}, \binits{A.}},
\oauthor{\bsnm{Guermond}, \binits{J.-L.}}:
Finite Elements I: Approximation and Interpolation.
Springer
(2021)
\end{botherref}
\endbibitem

\bibitem[\protect\citeauthoryear{Galarce~Marin}{2021}]{galarceThesis}
\begin{botherref}
\oauthor{\bsnm{Galarce~Marin}, \binits{F.}}:
Inverse problems in hemodynamics. fast estimation of blood flows from medical
  data. https://gitlab.com/felipe.galarce.m/mad/.
PhD thesis,
INRIA Paris \& Laboratoire Jacques-Louis Lions. Sorbonne Universit\'e
(2021)
\end{botherref}
\endbibitem

\bibitem[\protect\citeauthoryear{Galarce et~al.}{2021}]{GLM2021}
\begin{barticle}
\bauthor{\bsnm{Galarce}, \binits{F.}},
\bauthor{\bsnm{Lombardi}, \binits{D.}},
\bauthor{\bsnm{Mula}, \binits{O.}}:
\batitle{Reconstructing haemodynamics quantities of interest from {D}oppler
  ultrasound imaging}.
\bjtitle{Int. J. Numer. Meth. Biomedical Eng.}
(\byear{2021})
\doiurl{10.1002/cnm.3416}
\end{barticle}
\endbibitem

\bibitem[\protect\citeauthoryear{Galarce et~al.}{2023}]{galarce-etal-2023}
\begin{barticle}
\bauthor{\bsnm{Galarce}, \binits{F.}},
\bauthor{\bsnm{Tabelow}, \binits{K.}},
\bauthor{\bsnm{Polzehl}, \binits{J.}},
\bauthor{\bsnm{Panagiotis}, \binits{C.P.}},
\bauthor{\bsnm{Vavourakis}, \binits{V.}},
\bauthor{\bsnm{Lilaj}, \binits{L.}},
\bauthor{\bsnm{Sack}, \binits{I.}},
\bauthor{\bsnm{Caiazzo}, \binits{A.}}:
\batitle{Displacement and pressure reconstruction from magnetic resonance
  elastography images: Application to an in silico brain model.}
\bjtitle{SIAM Journal on Imaging Science}
(\byear{2023})
\doiurl{10.1137/22M149363X}
\end{barticle}
\endbibitem

\bibitem[\protect\citeauthoryear{Galarce et~al.}{2022}]{GLM2021_2}
\begin{botherref}
\oauthor{\bsnm{Galarce}, \binits{F.}},
\oauthor{\bsnm{Lombardi}, \binits{D.}},
\oauthor{\bsnm{Mula}, \binits{O.}}:
State estimation with model reduction and shape variability. application to
  biomedical problems.
{SIAM} Journal on Scientific Computing
\textbf{44}
(2022)
\doiurl{10.1137/21M1430480}
\end{botherref}
\endbibitem

\bibitem[\protect\citeauthoryear{Galarce and Pacheco}{2024}]{galarcePachecoBVS}
\begin{botherref}
\oauthor{\bsnm{Galarce}, \binits{F.}},
\oauthor{\bsnm{Pacheco}, \binits{D.}}:
Fully consistent lowest-order finite element methods for generalised stokes
  flows with variable viscosity.
ArXiv preprint
(2024)
\end{botherref}
\endbibitem

\bibitem[\protect\citeauthoryear{Balay et~al.}{2015}]{petsc}
\begin{botherref}
\oauthor{\bsnm{Balay}, \binits{S.}},
\oauthor{\bsnm{Abhyankar}, \binits{S.}},
\oauthor{\bsnm{M.D.}, \binits{A.}},
\oauthor{\bsnm{Brown}, \binits{J.}},
\oauthor{\bsnm{Brune}, \binits{P.}},
\oauthor{\bsnm{Buschelman}, \binits{K.}},
\oauthor{\bsnm{Dalcin}, \binits{L.}},
\oauthor{\bsnm{Eijkhout}, \binits{V.}},
\oauthor{\bsnm{Gropp}, \binits{W.D.}},
\oauthor{\bsnm{Kaushik}, \binits{D.}},
\oauthor{\bsnm{Knepley}, \binits{M.G.}},
\oauthor{\bsnm{McInnes}, \binits{L.C.}},
\oauthor{\bsnm{Rupp}, \binits{K.}},
\oauthor{\bsnm{Smith}, \binits{B.F.}},
\oauthor{\bsnm{Zampini}, \binits{S.}},
\oauthor{\bsnm{Zhang}, \binits{H.}}:
{PETS}c {W}eb page
(2015).
\url{http://www.mcs.anl.gov/petsc}
\end{botherref}
\endbibitem

\bibitem[\protect\citeauthoryear{Amestoy et~al.}{2000}]{AMESTOY2000501}
\begin{barticle}
\bauthor{\bsnm{Amestoy}, \binits{P.R.}},
\bauthor{\bsnm{Duff}, \binits{I.S.}},
\bauthor{\bsnm{L'Excellent}, \binits{J.-Y.}}:
\batitle{Multifrontal parallel distributed symmetric and unsymmetric solvers}.
\bjtitle{Computer Methods in Applied Mechanics and Engineering}
\bvolume{184}(\bissue{2}),
\bfpage{501}--\blpage{520}
(\byear{2000})
\doiurl{10.1016/S0045-7825(99)00242-X}
\end{barticle}
\endbibitem

\bibitem[\protect\citeauthoryear{Amestoy et~al.}{2001}]{mumps}
\begin{barticle}
\bauthor{\bsnm{Amestoy}, \binits{P.R.}},
\bauthor{\bsnm{Duff}, \binits{I.S.}},
\bauthor{\bsnm{L'Excellent}, \binits{J.-Y.}},
\bauthor{\bsnm{Koster}, \binits{J.}}:
\batitle{A fully asynchronous multifrontal solver using distributed dynamic
  scheduling}.
\bjtitle{SIAM Journal on Matrix Analysis and Applications}
\bvolume{23}(\bissue{1}),
\bfpage{15}--\blpage{41}
(\byear{2001})
\doiurl{10.1137/S0895479899358194}
{\href{https://arxiv.org/abs/https://doi.org/10.1137/S0895479899358194}{{https://doi.org/10.1137/S0895479899358194}}}
\end{barticle}
\endbibitem

\bibitem[\protect\citeauthoryear{Brezzi and Pitkaranta}{1984}]{brezzi1984}
\begin{botherref}
\oauthor{\bsnm{Brezzi}, \binits{F.}},
\oauthor{\bsnm{Pitkaranta}, \binits{J.}}:
On the stabilization of finite element approximations of the stokes equations.
Efficient Solutions of Elliptic Systems
(1984)
\end{botherref}
\endbibitem

\bibitem[\protect\citeauthoryear{Tezduyar}{1991}]{supg}
\begin{botherref}
\oauthor{\bsnm{Tezduyar}, \binits{T.E.}}:
Stabilized finite element formulations for incompressible flow computations
\textbf{28},
1--44
(1991)
\doiurl{10.1016/S0065-2156(08)70153-4}
\end{botherref}
\endbibitem

\bibitem[\protect\citeauthoryear{González et~al.}{2020}]{GONZALEZ20201009}
\begin{barticle}
\bauthor{\bsnm{González}, \binits{A.}},
\bauthor{\bsnm{Castillo}, \binits{E.}},
\bauthor{\bsnm{Cruchaga}, \binits{M.A.}}:
\batitle{Numerical verification of a non-residual orthogonal term-by-term
  stabilized finite element formulation for incompressible convective flow
  problems}.
\bjtitle{Computers \& Mathematics with Applications}
\bvolume{80}(\bissue{5}),
\bfpage{1009}--\blpage{1028}
(\byear{2020})
\doiurl{10.1016/j.camwa.2020.05.025}
\end{barticle}
\endbibitem

\bibitem[\protect\citeauthoryear{Guermond
  et~al.}{2005}]{overviewSolvingNSincomp}
\begin{barticle}
\bauthor{\bsnm{Guermond}, \binits{J.}},
\bauthor{\bsnm{Minev}, \binits{P.}},
\bauthor{\bsnm{Shen}, \binits{J.}}:
\batitle{An overview of projection methods for incompressible flows}.
\bjtitle{Computer Methods in Applied Mechanics and Engineering}
(\byear{2005})
\doiurl{10.1016/j.cma.2005.10.010}
\end{barticle}
\endbibitem

\bibitem[\protect\citeauthoryear{Brooks and Hughes}{1982}]{BROOKS1982199}
\begin{barticle}
\bauthor{\bsnm{Brooks}, \binits{A.N.}},
\bauthor{\bsnm{Hughes}, \binits{T.J.R.}}:
\batitle{Streamline upwind/{P}etrov-{G}alerkin formulations for convection
  dominated flows with particular emphasis on the incompressible navier-stokes
  equations}.
\bjtitle{Computer Methods in Applied Mechanics and Engineering}
\bvolume{32}(\bissue{1}),
\bfpage{199}--\blpage{259}
(\byear{1982})
\doiurl{10.1016/0045-7825(82)90071-8}
\end{barticle}
\endbibitem

\bibitem[\protect\citeauthoryear{Codina et~al.}{2017}]{RamonReview2017}
\begin{botherref}
\oauthor{\bsnm{Codina}, \binits{R.}},
\oauthor{\bsnm{Badia}, \binits{S.}},
\oauthor{\bsnm{Baiges}, \binits{J.}},
\oauthor{\bsnm{Principe}, \binits{J.}}:
Variational Multiscale Methods in Computational Fluid Dynamics. {E}ncyclopedia
  of Computational Mechanics Second Edition.
John Wiley \& Sons, Ltd
(2017).
\doiurl{10.1002/9781119176817.ecm2117} .
\url{https://onlinelibrary.wiley.com/doi/abs/10.1002/9781119176817.ecm2117}
\end{botherref}
\endbibitem

\bibitem[\protect\citeauthoryear{Alejo and Barrientos}{2009}]{alejo2009}
\begin{barticle}
\bauthor{\bsnm{Alejo}, \binits{B.}},
\bauthor{\bsnm{Barrientos}, \binits{A.}}:
\batitle{Model for yield stress of quartz pulps and copper tailings}.
\bjtitle{International Journal of Mineral Processing}
\bvolume{93}(\bissue{3-4}),
\bfpage{213}--\blpage{219}
(\byear{2009})
\doiurl{10.1016/j.minpro.2009.08.002}
\end{barticle}
\endbibitem

\bibitem[\protect\citeauthoryear{Centre}{2025}]{pipe}
\begin{botherref}
\oauthor{\bsnm{Centre}, \binits{A.S.}}:
Schedule 40 Steel Pipe.
\url{https://www.amardeepsteel.com/schedule-40-steel-pipe.html} [Accessed:
  2025-05-29]
(2025)
\end{botherref}
\endbibitem

\end{thebibliography}
\end{document}